\pdfoutput=1
\documentclass[11.5pt]{elsarticle}
\usepackage{amsmath,amsthm,amssymb} 
\usepackage{flexisym} 
\usepackage[pagewise]{lineno}
\usepackage{breqn}
\usepackage[english]{babel}
\usepackage{lmodern}
\usepackage{avant}
\usepackage{rotating}
\usepackage{adjustbox}
\usepackage{multirow}
\usepackage{booktabs}
\usepackage{appendix}
\usepackage{courier}
\usepackage{makecell}

\usepackage{subcaption}
\usepackage[T1]{fontenc}
\usepackage[latin9]{inputenc}
\usepackage[a4paper]{geometry}
\usepackage{mathtools}   
\usepackage{float}
\usepackage{stmaryrd}
\usepackage{setspace}
\usepackage{natbib}
\setcitestyle{round} 
\setcitestyle{authoryear} 
\PassOptionsToPackage{normalem}{ulem}
\usepackage{ulem}
\usepackage[unicode=true, bookmarks=true,bookmarksnumbered=false,bookmarksopen=false,
 breaklinks=false,pdfborder={0 0 0},pdfborderstyle={},backref=false,colorlinks=true]
 {hyperref}
\hypersetup{
 pdfborderstyle={},pdfborderstyle={},linkcolor=blue,urlcolor=blue,citecolor=blue}
\usepackage{units}
\allowdisplaybreaks

\makeatletter
\numberwithin{equation}{section}

\theoremstyle{definition}
\newtheorem{example}{Example}

\begin{document}

\begin{frontmatter}

\title{A robust statistical framework for cyber-vulnerability prioritisation under partial information in threat intelligence}

\author[1]{Mario Angelelli\corref{cor1}}
\ead{mario.angelelli@unisalento.it}
\cortext[cor1]{Corresponding Author}

\author[1]{Serena Arima}
\ead{serena.arima@unisalento.it}

\author[4]{Christian Catalano}
\ead{christian.catalano@uniba.it}

\author[1,5]{Enrico Ciavolino}
\ead{enrico.ciavolino@unisalento.it}

\affiliation[1]{
    organization={University of Salento}, 
    addressline={Edificio 5 - Complesso Studium 2000, Via di Valesio}, 
            city={Lecce},
            postcode={73100}, 
            country={Italy}
            }
\affiliation[4]{
    organization={University of Bari ``Aldo Moro''}, 
    addressline={via E. Orabona 4}, 
            city={Bari},
            postcode={70125}, 
            country={Italy}
}
\affiliation[5]{
    organization={WSB Merito University}, 
            city={Gdansk},
            country={Poland}
            }

\date{}

\begin{abstract} 
Proactive cyber-risk assessment is gaining momentum due to the wide range of sectors that can benefit from the prevention of cyber-incidents by preserving integrity, confidentiality, and the availability of data. The rising attention to cybersecurity also results from the increasing connectivity of cyber-physical systems, which generates multiple sources of uncertainty about emerging cyber-vulnerabilities.

This work introduces a robust statistical framework for quantitative and qualitative reasoning under uncertainty about cyber-vulnerabilities and their prioritisation. Specifically, we take advantage of mid-quantile regression to deal with ordinal risk assessments, and we compare it to current alternatives for cyber-risk ranking and graded responses. For this purpose, we identify a novel accuracy measure suited for rank invariance under partial knowledge of the whole set of existing vulnerabilities.

The model is tested on both simulated and real data from selected databases that support the evaluation, exploitation, or response to cyber-vulnerabilities in realistic contexts. Such datasets allow us to compare multiple models and accuracy measures, discussing the implications of partial knowledge about cyber-vulnerabilities on threat intelligence and decision-making in operational scenarios.
\end{abstract}


\end{frontmatter}

\section{\label{sec: Introduction} Introduction}

Cyber-vulnerabilities of devices, networks, or other 
information and communication technologies (ICTs) can generate system failures or pave the way for different types of cyber-attacks, including denial-of-service, malware injection, and data exfiltration. Social engineering can also enhance these incidents, while cascading effects in complex ICTs or systems-of-systems \citep{Fortino2020} can compromise or interrupt service supply, undermining the operational continuity of critical infrastructures. In turn, cyber-incidents lead to economic losses, safety risks, reputational damage, and violations of personal rights such as privacy, the right-to-be-anonymous, and the proper use of personal or sensitive data. The effect of these damages is not always measurable due to the intangible nature of social and reputational effects and the lack of high-quality data, which are often kept secret to prevent additional reputational issues \citep{Giudici2021}. 

New vulnerabilities emerge from the increasing number of connections between digital systems, which now include personal devices, Internet-of-Things (IoT) sensors, cloud computing or storage services, and even vehicles \citep{Barletta2023}, which represent access points to other information systems through privilege escalation. The latter amplifies the severity of cyber-vulnerabilities and represents a weakness when local access points may lead to violations of classified information at the national level, as in the case of public administration \citep{Catalano2021}. 

To prevent cyber-incidents, \emph{proactive} cyber-risk assessment keeps evolving through new methods, standards, approaches, and good practices aimed at informed decision-making in the management of cyber domains, in particular cyber-vulnerabilities. Currently, cyber-risk assessment standards are based on severity levels assessed by institutions, such as the National Institute of Standards and Technology (NIST) and national Computer Security Incident Response Teams (CSIRTs). Although NIST provides a harmonised approach to evaluating the general impact of a cyber-vulnerability, contextual factors (e.g., exposure to a vulnerable technology and its identifiability) may influence exploitability. Available information on these factors may affect the perceived likelihood of a cyber-attack exploiting a cyber-vulnerability, influencing both offensive and defensive interventions and resource usage. Such information is often stored in reserved reports, data collections, or expert evaluations that are not disclosed. In addition to this limited knowledge, multiple cyber-vulnerabilities can be relevant to individuals and organisations, which have to prioritise them to better allocate their cybersecurity (economic, temporal, and professional) resources based on accessible information and personal criteria.

These issues prompt a deeper analysis of the way risk about cyber-vulnerabilities is perceived and evaluated based on available information: this leads to the following research questions (RQs): 
\begin{itemize}
    \item[(RQ1)] How to \emph{assess} cyber-risk based on partial information on known vulnerabilities without relying on specific statistical properties (e.g., their distributional assumptions) that could hardly be verified?
    \item[(RQ2)] How to \emph{measure} the accuracy of such an assessment while also taking into account the presence of unknown vulnerabilities?
\end{itemize}

To answer these questions, we propose a new statistical framework to address the need for flexible and interpretable models relating to cyber-vulnerability assessment and their prioritisation, in this way supporting adaptive decision-making. Flexibility is required to allow different users to adapt the framework based on the information they have access to, e.g., by adding explanatory variables or considering different response variables based on their own ranking. Interpretability is needed to prompt appropriate interventions, e.g., counteractions to fix vulnerabilities or prevent their exploitation. 

This work focuses on vulnerabilities rather than actual incidents, which requires appropriate models to deal with the two types of uncertainty connected to the research questions in terms of both estimation procedures and accuracy measures. Specifically, to address RQ1, we adopt mid-quantile regression \citep{Geraci2022} as a means to provide robust estimates of ordinal (quantitative and qualitative) risk assessments of known cyber-vulnerabilities dependent on available information. Regarding RQ2, we introduce a new accuracy measure that meets an invariance requirement for cyber-vulnerability priority rankings with respect to unobserved or unknown vulnerabilities.

These proposals are tested on both simulated and real data; the former allow us to explore multiple scenarios and test the sensitivity of the assessment performance on hyperparameters and model assumptions, while the latter inform us on actual cyber-vulnerabilities, the extent to which they adhere to or deviate from parametric models, and the way the different methods perform under such deviations. We summarise the main contributions of this work as follows: 
\begin{itemize}
    \item The first methodological contribution is mid-quantile-based statistical models to work out qualitative variables with quantitative methods. This proposal allows for overcoming the dependence on statistical assumptions, enabling the prediction of both qualitative and quantitative priority measures. Along with robust quantile regression estimates, these models return conditional probability estimates for an ordinal response variable, so they may serve as a basis for novel probabilistic modelling of cyber-threat assessment and risk analysis relying on likelihood estimations associated with a given impact \citep{crotty2022cyber} if an appropriate set of explanatory variables is available. 
    \item As the focus of this work is on cyber-vulnerability prioritisation, the second theoretical contribution is the proposal of a new accuracy index for rank prediction. The definition of this index is grounded in the inherent uncertainty of unknown vulnerabilities. By relying on both simulated and real data, we can explore the properties of the new accuracy measures, in particular their ability to discriminate between different ranking models in terms of prediction accuracy, depending on hyperparameters (e.g., the number of priority levels) or deviations from widely adopted statistical assumptions. 
    \item Along with the methodological contributions, we carry out a data collection procedure to test our proposals, integrating information from multiple datasets, discussing the results in relation to recent studies, and pointing out implications in cyber-vulnerability prioritisation for research in threat intelligence.  
\end{itemize}
While the statistical approach presented here is flexible enough to include other threat sources, the data we consider in this work do not involve factors such as social engineering, insider threats, or physical effects (e.g., overload of ICT capacities). However, it is worth stressing that such factors may be as critical as cyber-vulnerabilities and may combine with them in the execution of a cyber-attack \citep{Catalano2022}.  

The paper is organised as follows: the notions of cybersecurity and cyber-vulnerabilities that are relevant for this work are described in Section \ref{sec: cyber-risk basic notions and data sources}, where we also present an overview of recent advances in related works and introduce the required preliminaries on the statistical models used in the paper. 
Our proposal is presented and motivated in Section \ref{sec: our proposal}, also discussing the appropriate index to assess performance and model comparison suited to our research questions in the cyber-risk domain. Section \ref{sec: data sources} describes the data sources that are used for the specification and validation of the proposed model. In Section \ref{sec: result}, following a descriptive analysis of the data, we summarise and comment on the results of simulations and the exploration of the real dataset in terms of prioritising cyber-vulnerabilities. After the discussion of the results in Section \ref{sec: discussion}, conclusions are drawn in Section \ref{sec: conclusion}, where we point out future work and applications of the present proposal.

\section{Related work}
\label{sec: cyber-risk basic notions and data sources}

Cyber-risk assessment is a well-recognised issue that plays a key role in different domains, e.g., the management of critical infrastructures \citep{Pate2018} and industrial sectors \citep{Corallo2020}. Cyber-physical systems and personal devices require adequate solutions to ensure data protection, and the diffusion of IoT is opening the way to new sources of cyber-risk \citep{Radanliev2018,Tsiknas2021}. A variety of cyber-risk models have been introduced to support risk assessment and prioritisation, but their effectiveness in operational scenarios is affected by domain-specific aspects and requires an appropriate trade-off between the assessment's validity and its usability for decision-making \citep{Pate2018}. 

\subsection{Cyber-risk assessment and modelling}
\label{subsec: related work on cyber-risk modelling}

The scope of the cyber-risk assessment should be clarified by first specifying the objective of the analysis (e.g., proactive prevention or forensic investigation), the object of the analysis, and, consequently, the methodology adopted. This work is focused on proactive prevention, where one should distinguish between cyber-vulnerability and cyber-incident: a vulnerability is an access point, but this does not necessarily entail a cyber-incident, that is, actual (intentional or not) damage to a digital system. This distinction is relevant for decision-makers, namely, cybersecurity experts and ICT managers, security operational centres, or national agencies. Cyber-incident analysis is fundamental to cyber-forensic activities. Still, the prevention of \emph{new} cyber-incidents in operational scenarios should use all fungible information to manage security resources better and take appropriate counteractions.

Each known cyber-vulnerability is uniquely identified by a Common Vulnerability Exposure (CVE) code. In the NIST classification, the CVE acts as a primary key to retrieving both the impacts in terms of CIA dimensions (confidentiality, integrity, and availability) and the severity assessment of relevant intrinsic characteristics of the vulnerability. Focusing on cyber-vulnerabilities as the object of our assessment, the standard approach to properly scoring emergent vulnerabilities is driven by the NIST's methodology \citep{Sharma2018,Jung2022}. 

In addition to such intrinsic features of cyber-vulnerabilities, other extrinsic factors affect cyber-risk and threats, in particular a technology's \emph{exposure}, which refers to the number of exposed hosts (devices or systems) where a given vulnerability, labelled by a CVE, has been recognised. Exposure concurs to define targets and feasible attacks along with \emph{exploits} and their cost; an exploit is defined as a software component, a process, or any human or physical resource that can be directly executed to perform a cyber-attack. In this work, we primarily deal with software exploits, but related work also addresses the role of interactions between malicious software and human factors in the definition of new attack techniques \citep{Tommasi2022}. We talk about a 0\emph{-day} when the vulnerability has not been disclosed before and there are no available solutions to patch it. 

Proactive defence aims at increasing resilience at the individual and network level (preventing criticalities), supporting efficient management of resources and ICT maintenance, and preserving individuals and community rights in cyber-space such as privacy, compliance with the General Data Protection Regulation (GDPR), and right-to-be-anonymous. In particular, proactive defence is needed to choose appropriate counteractions that mitigate the occurrence of cyber-incidents from cyber-vulnerabilities. There are several techniques to enhance cybersecurity, including vulnerability assessment, penetration testing, and static or dynamic analysis of applications. However, proactive defence is subject to bounded resources: time constraints, verification costs \citep{Srinidhi2015,Gao2022}, a specific effort for proprietary software, limits to automation, and contextual security analysis in highly connected systems. Therefore, accurate methods to support experts in risk assessment are a relevant premise for prioritising interventions and, hence, making better use of resources. In this regard, (semi-)automatic tools and applications based on AI, especially deep learning, are gaining increasing attention as practical support to detect malware \citep{Cui2018}. Unfortunately, they do not provide complete protection against malware attacks; in a recent study \citep{Catalano2022}, 
it was shown that classification based on convolutional neural networks could be deceived by masking malware with a goodware component to bypass automatic controls. This approach is called \emph{polymorphism} and is a software property often used in cyber guerrilla attacks \citep{Van2016}. Furthermore, \citet{Macas2023} conducted a detailed review and categorisation of cyber-attacks taking advantage of adversarial learning. On the other hand, these works outline potential counteractions to mitigate cyber-risks in relation to such applications of deep learning. Also, new approaches are being investigated to benefit from deep learning while overcoming some of its limitations, e.g., enhancing explainability \citep{Keshk2023,Sharma2023}. 

Moving to risk assessment methodologies and modelling, different research streams are investigated to support cybersecurity experts through different methodological or algorithmic techniques. Qualitative approaches supporting cyber-risk management are recommended in international standards, including risk matrices. However, the validity of such approaches is limited by methodological issues that can lead to inconsistencies, misleading interpretations, and a lack of focus on potential correlations among risk factors (see, e.g., \citet{crotty2022cyber} and references therein). 

On the other hand, partial information in the cybersecurity domain is a serious obstruction to quantitative analysis, which influences its limited adoption compared to qualitative or semi-qualitative methods based on risk matrices. In fact, limited data accessibility has been widely recognised as a relevant issue \citep{Giudici2021}, with an economic impact on estimates \citep{Anderson2013} and consequent effects on insurance \citep{Carfora2019}. Among the main factors leading to data scarcity or non-availability, we mention resource limitations for conducting vulnerability assessments and non-disclosure policies to avoid sharing confidential information on cyber-threats and reputational losses. These aspects should be considered along with the lack of harmonisation between different quantitative methodologies, which hinders the assessments' comparability \citep{crotty2022cyber,facchinetti2023network}. 

A central topic in quantitative risk analysis is the way the likelihood and impact of a cyber-incident are estimated. Probability estimation is subject to various uncertainty sources and limitations in different quantitative methods \citep{allodi2017security}, and available assessments provided by cybersecurity agencies should be integrated with external information. For example, several studies adopt the CVSS as a means to evaluate the probability of a cyber-vulnerability's exploitation leading to a cyber-attack; see, e.g., the references in \citet[p. 168207]{he2019unknown}. Similar approaches are questioned by other works, which suggest that CVSS alone does not directly link to a cyber-attack's likelihood; instead, the CVSS should be combined with external information regarding exploits and available resources in the black market \citep{allodi2014comparing}.  

A general approach to data-driven updates of probability distributions by combining different information sources about cyber-vulnerabilities is given by Bayesian statistics and related computational techniques. The Factor Analysis of Information Risk (FAIR) model is a prominent example based on a well-established information security risk ontology; FAIR allows evaluating risk through the specification of a class of prior distributions and Monte Carlo simulations \citep{crotty2022cyber}. Even in this case, the model's applicability is limited by the adherence of specific scenarios in the cyber-domain with the model's distribution assumptions, and recent works have tested and relaxed such assumptions \citep{wang2020bayesian}. Related to this work, network-based approaches have been applied to cyber-risk modelling in different ways, starting with network analysis of connected hosts \citep{gil2014genetic} and including knowledge graphs \citep{zhao2023survey} and Bayesian networks or machine learning (e.g., random forest) algorithms \citep{facchinetti2023network,Kia2024}. Knowledge graphs allow encoding semantic structures and have strict relations with cybersecurity ontologies \citep[Sec.2]{zhao2023survey}, providing practical support in knowledge retrieval, reporting, and analysis in combination with statistical or machine learning algorithms. Bayesian networks are a powerful approach to exploring causal relations or dependences, for example, in attack chains; furthermore, they are also used to enhance the integration of qualitative frameworks and regulatory aspects that can affect cyber-risk \citep{shin2015development}. Bayesian networks can be integrated with other techniques, including taxonomic models based on the frequency and magnitude of threats and losses, such as the FAIR model mentioned above \citep{wang2020bayesian}. Estimation techniques in Bayesian networks rely on distributional assumptions or the knowledge of distribution parameters, and they can be affected by uncertainty about the dependence structure connecting vulnerabilities, devices, and attacks. Therefore, even for this class of methods, deviations from distributional assumptions or a lack of information to identify the probabilistic or statistical models could undermine the validity of the approach, as current studies point out \citep{allodi2017security,woods2021sok,Kia2024}. 

Aiming at fostering automatic assessments and reducing subjective experts' bias, new supervised methods for cyber-risk prediction based on CVEs have been recently proposed, where natural language processing and topic detection help predict vulnerabilities' likelihood and impact \citep{Kia2024}. Motivated by the same need to infer the likelihood and impact of a cyber-vulnerability's exploitation, fuzzy logic has been considered too \citep{dondo2008vulnerability}. The role of uncertainty in the cyber-domain is also relevant for the development of fuzzy techniques applied to intrusion detection systems \citep{Javaheri2023}, game-theoretic modelling of allocation and sharing cyber-defence resources \citep{Gao2022}, copula-based risk modelling for time series analysis of cyber losses \citep{zangerle2023modelling}, and stochastic processes for evaluating the resilience of a system based on Markov chains \citep{Zhang2021}.

\subsection{Preliminaries on statistical models}
\label{subsec: methodologies}

In line with the research questions stated in the Introduction, here we focus on interpretable statistical modelling and recently proposed applications to promote proper cyber-risk assessment and cybersecurity analysis. Before discussing the two specific models addressed in this work in the cybersecurity domain, we briefly review the ordered logit (OrdLog) model as a benchmark for regression with ordinal responses \citep{McCullagh1980}. 

\subsubsection{Ordered logit model}
\label{subsubsec: ordered logit model}

The OrdLog model is a Generalised Linear Model (GLM) suited to cumulative probability distributions for ordinal responses conditioned on explanatory variables. GLMs have proven useful with count response data as a means to predict the number of intrusions \citep{leslie2018statistical} or other count data related to cyber-attacks. These statistical models can support testing the distributional assumptions underlying such count data. \citet{leslie2018statistical} stress some issues already mentioned above, namely, the subjectivity of vulnerability scoring systems and the issues posed by a qualitative, rather than quantitative, structure, the partial knowledge about existing vulnerabilities, and the dependence on the adopted technology. 

The OrdLog model is specified as follows: let $y_{1},\dots,y_{n}$ be a sample of $n$ ordinal responses, and $\mathbf{X}$ be a vector of explanatory variables (or regressors). The OrdLog model aims at describing the effect of regressors on the odds 
\begin{equation}
    \log\frac{P(y\leq h |\mathbf{X})}{P(y \ge h |\mathbf{X})}=\alpha_{h}-\beta\cdot\mathbf{X},\quad h_{1}\leq h_{2}\Leftrightarrow \alpha_{h_{1}}\leq \alpha_{h_{2}}
    \label{eq: ordered logit} 
\end{equation}
where $P(y\leq h |\mathbf{X})$ (respectively, $P(y\geq h |\mathbf{X})$) is the left (respectively, right) cumulative probability associated with the $h$-th level of the response and conditioned to the observed values $\mathbf{X}$. The fit procedure estimates the model parameters, which are the level-specific intercepts $\alpha_{h}$ and the  $\beta$ coefficients that quantify the effects of regressors on the log-odds. This formulation assumes that the proportional odds hypothesis, namely, the log-ratio of the odds on the left-hand side of (\ref{eq: ordered logit}), depends on the ordinal level $h$ only through the scale coefficient $\alpha_{h}$, which does not depend on the variables $\mathbf{X}$. 

Despite the wide applicability of ordered logit or probit, more general approaches can be envisaged to overcome limitations from the potential violation of model assumptions (in this case, the proportional odds hypothesis). Another motivation stimulating research for new methodologies to deal with ordinal responses is the reduced interpretability of parameter estimates of GLMs with respect to simpler linear regression. This aspect is relevant in operational scenarios, where decision-makers should be able to interpret and quantify the impact of an explanatory variable without assuming background knowledge of the underlying statistical model. For this reason, we briefly present a recent proposal regarding the use of a regression model with ordinal responses in cyber-risk assessment. 

\subsubsection{Rank transform in linear regression}
\label{subsubsec: rank transform in linear regression}

A recent approach in \cite{Giudici2021} involves a linear regression model (which we refer to as LinReg) for data regarding cyber-\emph{incidents} and is based on the rank transform of a $n$-dimensional ordinal variable $Y$ with $k$ levels, that is, the set of ranks for each observation with a given prescription to handle ties \citep{Iman1979}. Formally, we move from the ordinal response $Y$ to the rank-transformed variable $R(Y)$ defined by 
\begin{eqnarray} 
Y & \mapsto & R(Y)\in\left\{r_{1},r_{2},\dots,r_{k}\right\}, \quad \text{where }\nonumber \\ 
r_{1}=1, & \quad & r_{h+1}=r_{h}+\#Y^{(-1)}(\{h+1\}),\quad h\in\{1,\dots,k-1\}
    \label{eq: Giudici, rank transform}
\end{eqnarray}
and $\#Y^{(-1)}(\{h+1\})$ denotes the number of observations of $Y$ whose value is $h+1$. The fit of the regression model
\begin{equation} 
    r_{i}=\beta_{0}+\beta\cdot{\bf X}_{i}+\varepsilon_{i},\quad\varepsilon\sim\mathcal{N}(0,\sigma^{2}),\,i\in\{1,\dots,n\}
    \label{eq: Giudici, regression}
\end{equation}
where $\mathcal{N}(0,\sigma^{2})$ is the centred normal distribution with variance $\sigma^{2}$ estimated from the data, is compared to the \emph{Rank Graduation Accuracy} (RGA) \citep{Giudici2021}
\begin{equation} 
\mathrm{RGA}:=\sum_{i=1}^{n}\frac{n}{i}\cdot\left(\frac{1}{n\overline{y}}\cdot\sum_{j=1}^{i}y_{\hat{r}_{j}}-\frac{i}{n}\right)^{2}
\label{eq: RGA}
\end{equation}
where test data $y_{1}, \dots,y_{n}$ have mean $\overline{y}$ and are ranked using the estimated ranks $\hat{r}$ obtained by fitting (\ref{eq: Giudici, regression}).

As anticipated, the choice of model (\ref{eq: Giudici, rank transform})-(\ref{eq: Giudici, regression}) is argued to provide more interpretable results supporting decision-making with respect to GLMs. However, the use of linear regression with rank transform may not be suited to dealing with cyber-vulnerabilities; contrary to actual cyber-incidents, vulnerabilities are subject to the different types of uncertainty mentioned above, especially in the cyber-guerrilla context \citep{Van2016}. 

From a methodological perspective, this means that several assumptions underlying the linear regression model may not be fulfilled when dealing with cyber-vulnerabilities. In particular, linear models rely on the normality assumption for the residuals, which may not be met in networks of digital systems; in fact, evidence shows that some relevant features of data breach datasets are well described by heavy-tail distributions \citep{Edwards2016}. 
Even the homoscedasticity assumption may not be fulfilled, and class unbalancing could make the linear model more sensitive to this violation, while quantile regression does not assume homoscedasticity.

\subsubsection{Quantile regression: remarks for cyber-risk assessment}
\label{subsubsec: quantile regression and cyber-risk}

Both the OrdReg and the LinReg models rely on assumptions that may be unverifiable in real datasets: unbalanced classes, deviations from normality, and a lack of complete knowledge of the space of potential vulnerabilities (unknown ones or $0$-days) may reduce the effectiveness of the aforementioned regression methods. In the cyber-domain, such hypotheses may actually not be verifiable due to the already-mentioned confidentiality and restrictions on data sharing. For this reason, we consider distribution-free approaches to make the analysis more robust against violations of statistical assumptions and concentrate on quantile regression \citep{Koenker2001}. 

Let $Q_{\tau}:=\inf_{y}\{y:\,\tau\leq F(y))$ be the $\tau$-th quantile of a random variable with cumulative distribution function (CDF) $F$. Quantile regression estimates $Q_{\tau}$ conditioning on $p$ regressors $\mathbf{X}$ 
\begin{equation}
    Q_{\tau}(y_{i}|\mathbf{X}_{i},\beta) = \mathbf{X}_{i}^{\mathtt{T}}\cdot \beta(\tau),\quad i\in\{1,\dots,n\}.
    \label{eq: quantile regression model}
\end{equation}
Parameter estimates $\hat{\beta}(\tau)\in\mathbb{R}^{p}$ come from the minimization of the loss function \citep{Koenker2001}
\begin{eqnarray}
	\hat{\beta}(\tau) & := & \mathrm{argmin}_{\beta\in\mathbb{R}^{p}} \sum_{i=1}^{n} \varrho_{\tau}\left(y_{i}-\mathbf{X}_{i}^{\mathtt{T}}\cdot \beta\right), \nonumber \\
	\varrho_{\tau}(u) & := & u\cdot (\tau-\mathbb{I}(u<0)) 
	\label{eq: minimisation QAE}
\end{eqnarray}
where $\mathbb{I}(X)$ is the characteristic function of a subset $X\subseteq\mathbb{R}$. 

In addition to increased robustness against model misspecification, the choice of quantile regression leads to a new parameter $\tau$ that naturally relates to the notion of Value-at-Risk (VaR) (also see \cite{Radanliev2018,Carfora2019} for a discussion of VaR in the cybersecurity context), in line with the purposes of this work. 

Different estimates can arise from different choices of the quantile level, which lets us compare different rankings or prioritisations at different quantile levels by looking at parameters associated with regressors. However, this aspect may lead to ambiguities if it is not properly linked to risk evaluation and decision-making, e.g., when ranking the attributes represented by the regressors \citep{Angelelli2022b}. This leads us to consider quantile regression, where the response explicitly refers to a vulnerability's priority.  

\subsubsection{Mid-quantile regression}
\label{subsubsec: mid-quantile regression}

Dealing with an ordinal response, we have to extend the quantile regression approach to discrete variables; for this purpose, we take advantage of \emph{mid-quantile} (MidQR hereafter) regression methods. Recent work by \citet{Geraci2022} applies mid-quantile regression \citep{Parzen2004} to discrete data: starting with a random variable $Y$ described by a categorical distribution $Y\sim\mathrm{cat}(p_{h},1\leq h\leq k)$ with $k$ levels, we set 
\begin{equation} 
\pi_{1}=\frac{1}{2}\cdot p_{1},\quad \pi_{h}=\frac{1}{2}\cdot p_{h}+\sum_{\ell=1}^{h-1}p_{\ell},\quad h\in \{2,\dots,k\}
\end{equation} 
which represents the evaluation of the \emph{mid-cumulative distribution function} $G_{Y}(y)=p(Y\leq y) - \frac{1}{2} p(Y = y)$ for the values $y_{1}<y_{2}<\dots<y_{n}$. Introducing $\pi_{0}=0$, $\pi_{k+1}=1$, $y_{0}=y_{1}$, and $y_{k+1}=y_{k}$, we can define the \emph{mid-quantile function} as 
\begin{equation} 
     H_{Y}(p) = \int_{0}^{1}\sum_{h=0}^{k+1}\left((1-\gamma)\cdot y_{h}+\gamma\cdot y_{h+1}\right)\cdot \delta\left((1-\gamma)\cdot\pi_{h}+\gamma\cdot\pi_{h+1}-p\right) d\gamma
\end{equation}
where $\delta(\cdot)$ is the Dirac distribution.
Setting $F(y):=p(Y\leq y)$ as before, estimators for unconditioned MidQR are obtained naturally, i.e., by the substitution of the estimates in the expression of the mid-quantile function. Such estimators enjoy good asymptotic consistency and normality for the sampling distribution; see \citet{Ma2011,Geraci2022}, and references therein. 

For a given link function $h(\cdot)$, we can consider a conditional mid-quantile function $H_{h(Y)|\mathbf{X}}(p) = \mathbf{X}^{\mathtt{T}}\cdot\beta(p)$ and estimate $\hat{G}_{Y|\mathbf{X}}(y|\mathbf{x})$ from samples $(\mathbf{x}_i,y_i)$, $i\in\{1,\dots,n\}$, through a non-parametric estimator that can encompass both continuous and discrete regressors \citep{Li2008}: 
\begin{eqnarray}
\hat{G}_{Y|\mathbf{X}}(y|\mathbf{x}) & = & \hat{F}_{Y|\mathbf{X}}(y|\mathbf{x}) - \frac{1}{2}\cdot \hat{m}_{Y|\mathbf{X}}(y|\mathbf{x}),
\nonumber \\ 
\hat{F}_{Y|\mathbf{X}}(y|\mathbf{x}) & = & \frac{n^{-1}\cdot\sum_{i=1}^{n}\mathbb{I}(y_{i}\leq y)K_{\lambda}(\mathbf{X}_{i},\mathbf{x})}{\hat{\delta}_{\mathbf{X}}(\mathbf{x})},\nonumber \\ 
\hat{m}_{Y|\mathbf{X}}(z_{j}|\mathbf{x}) & = & \hat{F}_{Y|\mathbf{X}}(z_{j}|\mathbf{x})-\hat{F}_{Y|\mathbf{X}}(z_{j-1}|\mathbf{x})
\end{eqnarray}
where $K_{\lambda}(\mathbf{X}_{i},\mathbf{x})$ is a kernel function with bandwidth $\lambda$, $\hat{\delta}_{\mathbf{X}}(\mathbf{x})$ is the kernel estimator of the marginal density of the explanatory variables $\mathbf{X}$, and $z_{1}<z_{2}<\dots<z_{k}$ are the distinct values taken by the observations $\{y_{1},\dots,y_{n}\}$ in the sample. In this way, we can obtain $\hat{G}_{Y|\mathbf{X}}(y|\mathbf{x})=\hat{F}_{Y|\mathbf{X}}(y|\mathbf{x})-\frac{1}{2}\cdot \hat{m}_{Y|\mathbf{X}}(y|\mathbf{x})$.
Estimates of coefficients $\beta$ follow from the minimization of the following quadratic loss function
\begin{equation}
    \mathrm{arg\,min}\psi_{n}(\beta;p),\quad \psi_{n}(\beta;p):=n^{-1}\cdot \sum_{i=1}^{n}\left(p-\hat{G}_{Y|\mathbf{X}}(h^{-1}(\mathbf{X}_{i}^{\mathtt{T}}\cdot\beta)\right)^2.
\end{equation}
The estimation and fitting procedures can be carried out using the {\tt R} package {\tt Qtools} developed by \citet{Geraci2022}.

\section{Contribution and proposed methodology}
\label{sec: our proposal}

The previous discussion points out the need to facilitate the transfer of qualitative structures and assessments into quantitative models, as both have practical advantages and limitations. Qualitative assessments are widely adopted in standards and guidelines and allow encoding experts' evaluations even when sufficient data for quantitative analyses are not available; on the other hand, they may give rise to inconsistencies and embed subjective factors or biases, especially in the assessment of probabilities related to cyber-events \citep{deSmidt2018perceptions}. Quantitative methods enhance the assessments' accuracy and reduce ambiguity, but their implementation requires sensitive information or confidential data that are generally not available. Furthermore, the validity of those methods may rely on distributional assumptions or the knowledge of parameters or dependencies, which may be limited for the same reasons. 

A way to combine the two approaches is to adopt quantitative models to analyse ordinal assessments of qualitative variables; specifically, mid-quantile methods involve fitting (mid-)conditional distribution functions for cyber-vulnerability priority levels based on available information, so we can convert CVSS qualitative information, in conjunction with other relevant risk factors  \citep{allodi2014comparing}, into probabilistic models. Starting with an ordinal response variable, we can also move from cyber-vulnerabilities' priority to ranking, enabling the comparison of different methodologies such as the LinReg model mentioned above. The non-parametric approach that we adopt avoids methodological issues that could compromise the validity of the analysis, making the estimated probability usable in multiple settings. Finally, an appropriate accuracy index is proposed to enhance the compatibility of ranking predictions with the original ordinal structure and the uncertainty related to unknown cyber-vulnerabilities.

\subsection{Estimation: MidQR for robust cyber-vulnerability assessment} 

For our purposes, MidQR is used to provide estimates of the conditional quantile given a set of regressors that includes both intrinsic vulnerability characteristics and external variables (exposure and exploit availability), with a qualitative priority assessment as our ordinal response variable. In addition to quantile estimates, we are interested in the mid-cumulative distribution function that describes the conditional probability of priority levels, as it helps to identify where a lack of complete information may have an effect. Such a conditional distribution concerns the quantity 
\begin{equation}
    F_{Y|\mathbf{X}}(Y\leq y|\mathbf{x})=\frac{P(Y\leq y\wedge \mathbf{X}=\mathbf{x})}{P(\mathbf{X}=\mathbf{x})}
    \label{eq: balance impact and rarity}
\end{equation}
where we focus on regressors $\mathbf{X}$ with a non-zero probability mass. The quantity (\ref{eq: balance impact and rarity}) can be seen as a balance of the joint occurrence of a given impact level with cyber-vulnerability features ($P(Y\leq y\wedge \mathbf{X}=\mathbf{x})$) and the features' likelihood ($P(\mathbf{X}=\mathbf{x})$). The different forms of uncertainty mentioned in the Introduction, such as underreported vulnerabilities, affect the evaluation of (\ref{eq: balance impact and rarity}) starting from the measurements $\mathbf{x}$, as we have limited knowledge of the sample space due to unknown vulnerabilities. 

As a subsequent step, the resulting estimates are used to predict the priority level of new vulnerabilities at a given quantile level and, then, prioritise them. This last step should enjoy some invariance properties for the predicted values to mitigate the effect of the aforementioned uncertainty on the ranking accuracy. This requirement has a practical effect in regression models dealing with both estimated ranking (LinReg) and, more generally, distributions of ordinal variables (such as MidQR). In the scope of this work, the performance index we introduce in the next section complements the estimation phase by taking into account the effects of partial knowledge about vulnerabilities on rankings. 

Experts' subjectivity in the assessment of regressors extracted from the attack vector is another source of uncertainty \citep{Kia2024}. Even if this work does not involve measurement errors for the explanatory variables $\mathbf{X}$ in the regression models, we point out that Bayesian methods are a viable approach to dealing with a mixture of experts and grouping multiple regression models in the context of cyber-vulnerability assessment \citep{Angelelli2022a}.

\subsection{A new performance index for cyber-risk prediction under uncertainty}
\label{subsec: a new performance index for cyber-risk estimation}

The uncertainty about the sample spaces, with consequent effects on the estimation of the priority assessment, is a major driver that prompts our research for a new approach to evaluating the accuracy of the assessment. 

Specifically, the use of quantitative values in (\ref{eq: RGA}) should take into account the nature of the variables in the model. The evaluation of (\ref{eq: RGA}) assumes an algebraic structure, formally, the semiring $(\mathbb{N},+,\cdot,0,1)$ of natural numbers for rankings or the ordered field $(\mathbb{R},+,\cdot,0,1)$ for regression, which is not necessarily linked to the original ordinal variables assessing the priority of a cyber-vulnerability. This algebraic structure is an artefact suited to the regression model and, hence, to the estimated variables (let them be the rank transform or the mid-quantile); the only effect derived from the ordinal variables is the order defining the summands in (\ref{eq: RGA}). It is worth noting that a similar observation also applies in other frameworks for uncertainty modelling, e.g., when dealing with structural representations of epistemic uncertainty in data-driven initiatives \citep{Angelelli2024}.  

Motivated by these considerations, we introduce a novel prediction accuracy index to accommodate the characteristics of cyber-vulnerability data. We consider a \emph{reverse} RGA index defined as $\mathrm{RGA}(r_{\mathrm{tr}},r_{\mathrm{est}})$, namely, we exchange the roles of the estimated $r_{\mathrm{est}}$ and the ``true'' $r_{\mathrm{tr}}$ rankings. We refer to such an index as the Agreement of Grounded Rankings (AGR) to stress the focus on the reference frame in the ranking, namely, the order structure and the limited knowledge of the set of cyber-vulnerabilities to be ranked.  

To better appreciate the need for appropriate use of the RGA index for unconventional cyber-risk assessment, we consider the case of sub-sampling, i.e., known subsets of an unknown family of cyber-vulnerabilities. This emulates the partial knowledge available due to $0$-days. 

\begin{example}
\label{exa: non-invariance}
We can consider the following $5$-dimensional rank vectors: 
\begin{equation}
c_{\mathrm{est}}:=
(1,3,2,2.9,10), \quad 
c_{\mathrm{tr},1}:= (1,3,2,2,9), \quad 
c_{\mathrm{tr},2}:= (1,5,3,3,7)
\label{eq: example non-invariance}
 \end{equation} 
where $c_{\mathrm{est}}$ derives from a given estimation procedure, while $c_{\mathrm{tr},u}$, $u\in\{1,2\}$, are two ``true'' rankings obtained from different knowledge 
about the state of a digital system and its sample space. Although they are different, the rankings $c_{\mathrm{tr},1}$ and $c_{\mathrm{tr},2}$ are consistent with the same attribution of ordinal levels: for the sake of concreteness, we can assume that the components of both $c_{\mathrm{tr},1}$ and $c_{\mathrm{tr},2}$ are generated by ranking the same ordinal assessment (``$10$'',``$6$'',``$8$'',``$8$'',``$3$''), where priority levels are ordered from  ``$10$'' to ``$1$''. In this case, the differences between $c_{\mathrm{tr},1}$ and $c_{\mathrm{tr},2}$ can arise from the existence of other elements in the two ranked sample spaces beyond those associated with the components of $c_{\mathrm{tr},1}$ and $c_{\mathrm{tr},2}$. The evaluation of $\mathrm{RGA}(y_{\mathrm{est}}, y_{\mathrm{tr},u})$ for $u\in\{1,2\}$ following the definition (\ref{eq: RGA}) does not satisfy invariance under changes in rankings that are generated by the same ordinal assessment. Indeed, we have 
\begin{equation}
\mathrm{RGA}(c_{\mathrm{est}},c_{\mathrm{tr},1})=
0.5161 \neq
0.3232 = \mathrm{RGA}(c_{\mathrm{est}},c_{\mathrm{tr},2}). 
\end{equation}
On the other hand, we find 
\begin{equation} 
\mathrm{AGR}(c_{\mathrm{est}},c_{\mathrm{tr},1})=\mathrm{RGA}(c_{\mathrm{tr},1},c_{\mathrm{est}})=
0.5272=\mathrm{RGA}(c_{\mathrm{tr},2},c_{\mathrm{est}})=\mathrm{AGR}(c_{\mathrm{est}},c_{\mathrm{tr},2}).
\end{equation}
It is clear that the latter equality holds for all the choices of $c_{\mathrm{est}},c_{\mathrm{tr,1}},c_{\mathrm{tr},2}$.
\end{example}

This shows that the AGR index resolves the lack of invariance under sub-sampling in RGA. The favourable invariance of the AGR index under rank transformations that are compatible with the same underlying ordinal assessment is in line with Luce's axiom of Independence of Irrelevant Alternatives \citep{Luce2005}, while some algebraic conditions related to this type of symmetry have been discussed in reasoning under uncertainty \citep{Angelelli2024}. 
Practically, this invariance is required when dealing with partial information about the space of potential cyber-vulnerabilities, which is the general situation faced by a decision-maker due to the occurrence of unknown vulnerabilities not exploited yet, $0$-days, and \emph{unconventional} cyber-attacks \citep{Van2016,Tommasi2022}.

\section{Data sources}
\label{sec: data sources}

\subsection{Databases}
\label{subsec: databases}

\begin{figure}[H]
    \centering
    \includegraphics[width=\textwidth]{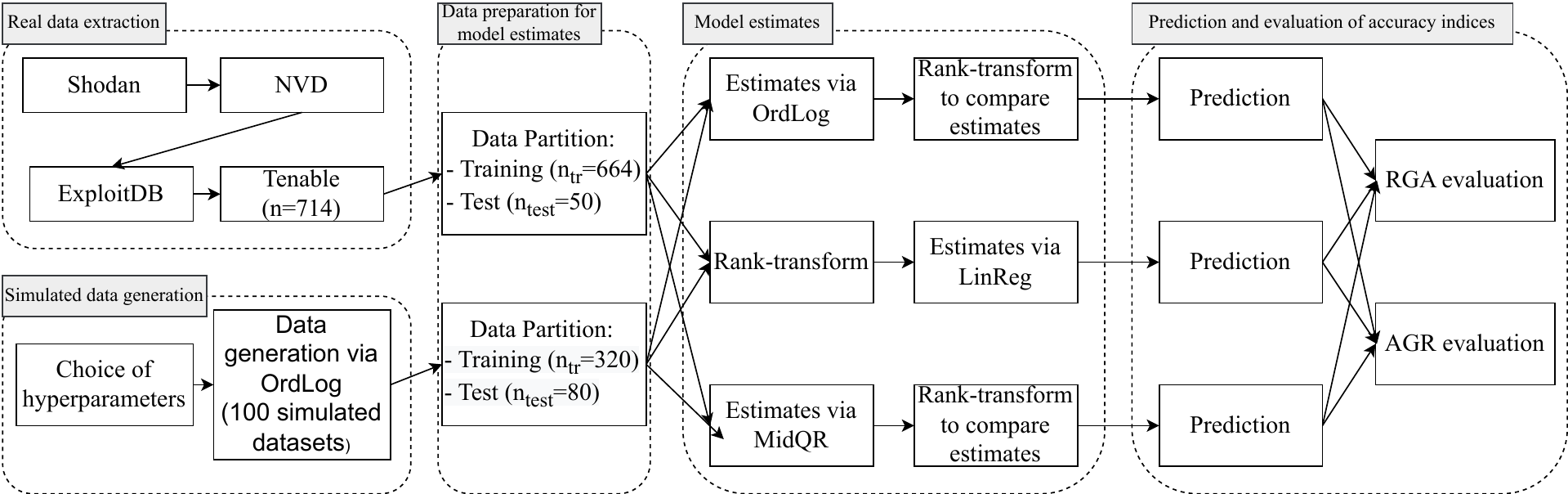}
    \caption{Graphical description of the experiments to validate the efficiency of mid-quantile regression for priority estimates and AGR as an accuracy index of predicted risk levels.}
    \label{fig: flowchart}
\end{figure}

Several databases can be used to assess the cybersecurity of a digital system. Among the most widely used by practitioners are the following ones:  
\begin{itemize}
    \item the National Vulnerability Database (NVD) includes assessments of vulnerabilities' severity by the NIST in terms of data impact dimensions (Confidentiality, Integrity, and Availability) and three additional technical features describing the accessibility prompted by the cyber-vulnerability, namely, Access Vector (AV), Access Complexity (AC), and Authentication (Au). The severity assessments of these six components compose the \emph{attack vector}\footnote{\url{https://nvd.nist.gov/vuln/search}}.
    \item The CSIRT database\footnote{\url{https://www.csirt.gov.it/contenuti/}} reports relevant updates on vulnerabilities in line with the evaluation by NIST. Such reports are communicated by the Italian CSIRT, which is established within the National Cybersecurity Agency. 
    \item The Shodan database\footnote{\url{https://exposure.shodan.io}} reports exposed hosts or IP addresses affected by known vulnerabilities, which may represent a relevant driver for attackers' intervention. The Shodan database can be queried by specifying a CVE and the country of the exposed hosts. Data are collected by the Shodan monitor platform by combining different techniques, such as crawling, IP lookups, and metadata analysis. 
    \item Reported exploits for CVEs can be extracted from ExploitDB\footnote{\url{https://www.exploit-db.com/}}. Information about exploits can be further refined from VulnDB\footnote{\url{https://vuldb.com/}}, a database that collects information on the price range of exploits associated with a CVE. The fields extracted from VulnDB include the 0-day price range, the price at the time of querying, and the exploitability. 
    \item Tenable's\footnote{\url{https://www.tenable.com/cve/search}} assessment \emph{interprets} CVSSs and assigns an ordinal risk priority through threat and vulnerability analysis. It contains qualitative risk information in Tenable's Vulnerability Priority Rating (VPR) assessment, which is obtained through machine learning algorithms that process information collected from the dark web, social media, code repositories, and reports. This index is the result of a threat intelligence activity that incorporates exploits' code maturity and extracts features to monitor the impact of a cyber-vulnerability in terms of actual and predicted threats\footnote{For more details on the VPR, we refer to \url{https://www.tenable.com/blog/what-is-vpr-and-how-is-it-different-from-cvss}}. 
\end{itemize}
For all these databases, we prepared Python scripts in order to extract the required data through APIs automatically: 
\begin{itemize} 
\item We started by selecting vulnerabilities identified in Italy through Shodan to obtain a base set of CVEs. Then, the \texttt{shodan} API was used to extract the exposure data. 
\item Subsequently, the scripts were adapted to extract the attack vectors associated with these CVEs from the NVD database through a request that returned a JSON file, which was inspected to get the CVSS scores. 
\item Then, we checked the availability of the exploits from ExploitDB and VulnDB. For ExploitDB, we used CVE Searchsploit \citep{cve_searchsploit} to obtain the exploits for the selected CVEs. 
\item In conclusion, a dedicated script was used to obtain Tenable's VPR assessment of the CVEs under consideration; even in this case, we collected these data by inspecting the output of a request for the selected CVEs.   
\end{itemize}
Running these Python scripts, the final dataset for model validation consists of $n=714$ units. This data extraction procedure is graphically depicted in Figure \ref{fig: flowchart} as a component of the overall analysis blue to validate the proposal and investigate its scope of applicability.

\subsection{Data description}
\label{subsec: data description}
The above data manipulation procedure leads to a dataset with the following variables: 
\begin{enumerate}
    \item Components of the attack vector obtained from the NIST vulnerability assessment constitute ordinal regressors.
    \item Exposure is a numerical variable that counts exposed hosts, but the variety of such count data lets us consider a continuous approximation of this variable. 
    \item For each CVE, the existence or absence of an exploit is encoded in a dichotomic variable.
    \item Tenable's priority rating is the ordinal response (dependent variable) that is linked to the previous explanatory variables through MidQR. 
\end{enumerate}
For the present investigation, we selected $p=7$ explanatory variables returned by the procedure described above, whose interpretation is summarised in Table \ref{tab: data structure}. 

\begin{table}[H]
\centering
\caption{Main attributes of the variables and their interpretation for statistical modelling. For each set of variables, the data source is provided in the leftmost column. The quantification for the ordinal assessments of the components  $X_{\mathrm{C}},X_{\mathrm{I}},X_{\mathrm{A}},X_{\mathrm{AV}},X_{\mathrm{AC}}$ of the attack vector (rightmost column) are provided by NVD experts.} 
\resizebox{\textwidth}{!}{%
\begin{tabular}{|c|l|l|l|l|}
\hline
\textbf{Source} &
  \multicolumn{1}{c|}{\textbf{Variables}} &
  \multicolumn{1}{c|}{\textbf{Type}} &
  \multicolumn{1}{c|}{\textbf{Interpretation}} &
  \multicolumn{1}{c|}{\textbf{Values}} \\ \hline
\multirow{6}{*}{\textbf{NIST}} &
  $X_{\mathrm{C}}$ &
  \multirow{6}{*}{\begin{tabular}[c]{@{}l@{}} Qualitative \\ Ordinal\end{tabular}} &
  Severity for Confidentiality &
  \multirow{3}{*}{\begin{tabular}[c]{@{}l@{}}$\bullet$ ``none: $0$''\\ $\bullet$ ``partial: $0.275$''\\ $\bullet$ ``complete: $0.660$''\end{tabular}} \\ \cline{2-2} \cline{4-4}
 &
  $X_{\mathrm{I}}$ &
   &
  Severity for Integrity &
   \\ \cline{2-2} \cline{4-4}
 &
  $X_{\mathrm{A}}$ &
   &
  Severity for Availability &
   \\ \cline{2-2} \cline{4-5} 
 &
  $X_{\mathrm{AV}}$ &
   &
  \begin{tabular}[c]{@{}l@{}}Type and severity \\ of the access vector\end{tabular} &
  \begin{tabular}[c]{@{}l@{}}$\bullet$ ``Requires local access: $0.395$\\ $\bullet$ ``Local Network accessible:  $0.646$'' \\ $\bullet$ ``Network accessible: $1$''\end{tabular} \\ \cline{2-2} \cline{4-5} 
 &
  $X_{\mathrm{AC}}$ &
   &
  \begin{tabular}[c]{@{}l@{}}Type and severity \\ of access complexity\end{tabular} &
  \begin{tabular}[c]{@{}l@{}}$\bullet$ ``high: $0.35$ \\ $\bullet$ ``medium: $0.61$'' \\ $\bullet$ ``low: $0.71$''\end{tabular} \\ 
 \hline
\textbf{Shodan} &
  $N_{\mathrm{exp}}$ &
  Count data &
  Number &
  Integers \\ \hline
\textbf{ExploitDB} &
  $q_{\mathrm{expl}}$ &
  Binary
  &
  Existence (Boolean)
  &
  $\{0,1\}$ (dichotomic)
  \\ \hline
\textbf{Tenable} &
  $Y$ &
  \begin{tabular}[c]{@{}l@{}} Qualitative \\ Ordinal\end{tabular} &
  \begin{tabular}[c]{@{}l@{}} Priority rating following \\ threat/vulnerability analysis\end{tabular} &
  \begin{tabular}[c]{@{}l@{}}``Low''\\ ``Medium'', \\ ``High'', \\ ``Critical''\end{tabular} \\ \hline
\end{tabular}%
}
\label{tab: data structure}
\end{table}

\section{\label{sec: result} Experiments and results}

\subsection{Descriptive analysis of the dataset} \label{subsec: descriptive analysis}
Data extracted from the databases described in Section \ref{sec: data sources} select $n=714$ cyber-vulnerabilities in Italy. The time span of the CVEs is 1999-2021. We concentrate on a single country to take into account local (country-wise) factors that could generate differences in cyber-risk and threat analyses  \citep{crotty2022cyber} and carry out the analysis within a known context. In our study, this choice may help to control contextual covariates that are not involved in this analysis, e.g., regulatory aspects and governance factors affecting both technological adoption and cyber-threats at a national level. We emphasise that this choice can be customised for other countries or extended on a cross-national scale based on the specific research design and assessment objectives. 

Regarding the time span, while the attack vector's components are intrinsic and, hence, do not change with time, the VPR and exposure are dynamically monitored and adapted, so they reflect the current state of the vulnerability within its limited life-cycle, also considering technology updating and cyber-vulnerability patching or fixing. By taking the exploit variable as dichotomic (existence or absence), we overcome potential temporal effects related to the number of exploits, which fall beyond the scope of the present analysis. However, we stress that the aforementioned regression models can capture temporal factors through relations between independent variables (in particular, exposure and exploit availability) and the dependent response (Tenable's VPR assessment). A dedicated study of these relations could align with and complement time-series analysis of the information in CVE scores and descriptions \citep{Kia2024}. 

\begin{figure}[H]
\begin{center}
\begin{subfigure}{.45\textwidth}
    \centering
    \includegraphics[width=\textwidth]{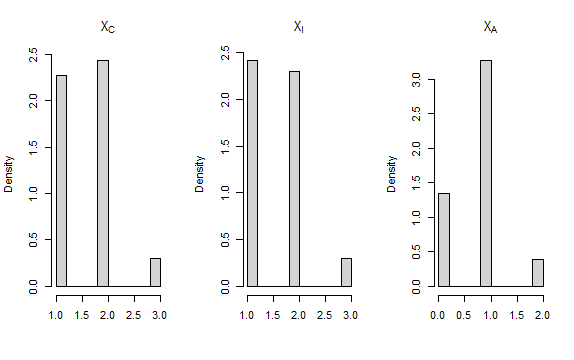}
    \caption{Distribution of impact dimension levels.}
  \label{fig: plot impacts}
\end{subfigure}%
\hfill
\begin{subfigure}{.45\textwidth}
    \centering
    \includegraphics[width=\textwidth]{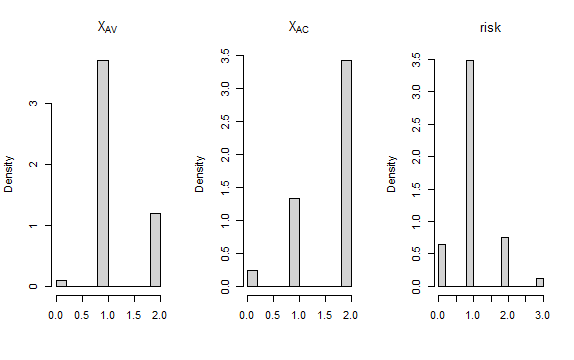}
    \caption{Distribution of features (AV, AC) and VPR.}
  \label{fig: plot features and risk script}
\end{subfigure}%
\caption{Distribution of levels of variables from the cyber-vulnerability dataset.}
\label{fig: plot distributions of regressors and risk factors}
\end{center}
\end{figure}

We note that each variable in the attack vector is characterised by manifest unbalancing among the different levels, as shown in Figures \ref{fig: plot impacts}-\ref{fig: plot features and risk script}. 
When the 
response in a regression model is well approximated by a continuous variable, then unbalancing could make linear regression more sensitive to deviations from homoscedasticity; hence, quantile regression could be favourable. 
This is the case when the exposure of vulnerable hosts is related to intrinsic features of the vulnerabilities \citep{Angelelli2022b}: it is easily checked from the QQ-plots in Figures \ref{fig: qq-plot residuals, log-integer}-\ref{fig: qq-plot free residuals, log-integer} that the residuals of the exposure $N_{\mathrm{exp}}$ and its log-transform $10\cdot \log_{10}(1+N_{\mathrm{exp}})$, considered as responses in a linear model with regressors $(X_{\mathrm{C}},X_{\mathrm{I}},X_{\mathrm{A}},X_{\mathrm{AV}},X_{\mathrm{AC}})$, show strong deviations from normality.

\begin{figure}[H]
\begin{subfigure}[H]{0.45\textwidth}
 \centering
    \includegraphics[width=1.1\textwidth]{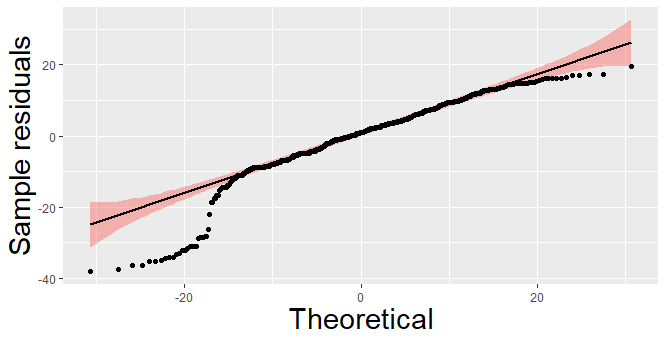}
\caption{Regressors $(X_{\mathrm{C}},X_{\mathrm{I}},X_{\mathrm{A}},X_{\mathrm{AV}},X_{\mathrm{AC}})$.}
\label{fig: qq-plot residuals, log-integer}
\end{subfigure}
\hfill
\begin{subfigure}[H]{0.45\textwidth}
\centering
\includegraphics[width=1.1\textwidth]{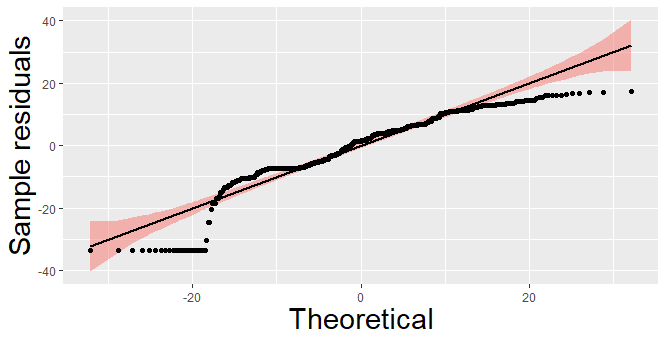}
\caption{Free model.}
  \label{fig: qq-plot free residuals, log-integer}
\end{subfigure}
\caption{QQ-plots of the theoretical (normal) quantiles compared to the empirical quantiles of residuals of $y=10\cdot \log_{10}(1+N_{\mathrm{exp}})$ derived from the exposure $N_{\mathrm{exp}}$ of cyber-vulnerabilities.}
\end{figure}

This remark also entails that linear regression would not fit the distribution assumptions when a proxy of cyber-risk, such as exposure, is used as the response. 
We also note that even the residuals of the ``free model'', i.e., the QQ-plot of the exposure $N_{\mathrm{exp}}$ itself, violate the normality assumption (see Figure \ref{fig: qq-plot free residuals, log-integer}). The use of the transform $N_{\mathrm{exp}}\mapsto 10\cdot \log_{10}(1+N_{\mathrm{exp}})$ in the previous QQ-plots slightly reduces the deviation from normality; more importantly, it highlights multimodality in the distribution of exposure, as it is manifest in the histograms depicted in Figures \ref{fig: histogram, integer}-\ref{fig: histogram, log-integer}.
\begin{figure}[H]
\begin{subfigure}{.5\textwidth}
    \centering
    \includegraphics[width=\textwidth]{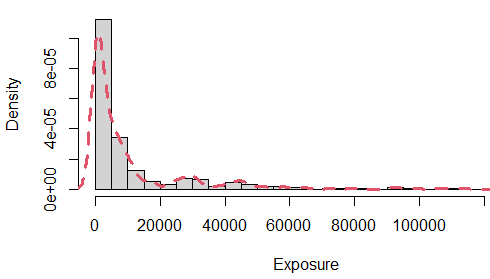}
    \caption{$y=N_{\mathrm{exp}}$.}
  \label{fig: histogram, integer}
\end{subfigure}%
\hfill
\begin{subfigure}{.5\textwidth}
    \centering
    \includegraphics[width=\textwidth]{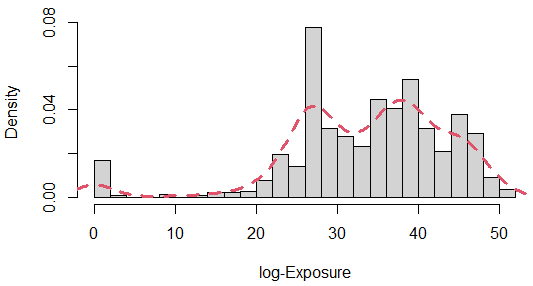}
    \caption{$y=10\cdot \log_{10}(1+N_{\mathrm{exp}})$.}
  \label{fig: histogram, log-integer}
\end{subfigure}%
\caption{Histograms for the empirical distributions of exposure $N_{\mathrm{exp}}$ compared to $10\cdot \log_{10}(1+N_{\mathrm{exp}})$. The corresponding continuous approximations (red dashed lines) highlight multimodality.}
\end{figure}

This suggests the need to go beyond linear models for an appropriate description of the external characteristics of cyber-vulnerabilities, starting from their intrinsic (attack vector) and extrinsic (exposure, exploits) features as regressors.

\subsection{Rankings and mid-quantile regression} 
\label{subsec: ranking and mid-quantile regression}

\subsubsection{Simulation study} 
\label{subsubsec: simulation specification}
Contrary to real dataset analysis, in this simulation study, we can control the data generation mechanism, so we can compare both estimation and accuracy measurement in relation to the data-generating model (OrdLog). Furthermore, we can conduct different tests to evaluate the models' performance at varying hyperparameters, in particular the number of ordinal levels in the response variable and the randomness of the probabilities in the OrdLog model. 

We start by specifying the preliminary simulation study to provide a general comparative analysis between the model presented in \citep{Giudici2021} and the MidQR. 
\begin{itemize}
    \item We used $n_{tr}= 320$ units for training and $n_{test}=80$ units for testing the accuracy performance of the models. We started with a response variable having $k = 4$ levels, in line with Tenable's priority rating that is used in the analysis of real data. However, we also tested $k\in\{3,6,8\}$ to evaluate the behaviour and performance of the different models when the number of levels of the response variable changes. 
    \item Two continuous and two factor explanatory variables were considered, each of the latter having three categories. This induced $P:=2+2\cdot(3-1)=6$ regressors after moving to ANOVA variables. 
    \item Following the generation of the so-specified variables, we considered the parameters $\alpha_{h}$, $h\in\{1,\dots,k-1\}$ and $\beta_{p}$, $p\in\{1,\dots,P\}$  to obtain the corresponding probabilities based on the ordered logit model (\ref{eq: ordered logit}). 
    \item This scheme was iterated to obtain $n_{iter}=100$ samples of the response variable $Y$. 
    \end{itemize}
In this way, we got the coefficient estimates and the mean, over the simulation runs, of the standard error (SE) estimates for each coefficient. For MidQR, we adapted a function in \texttt{Qtools} to overcome computational issues in the estimation of the conditional (mid-)CDF, which involves the kernel method 
based on \citep{Li2013}. 
Specifically, we acted on the estimated covariance matrix of the coefficients to make its computation compatible with cases where the quantile level lies outside the range of the sample mid-CDF. However, the outcomes of this procedure, which is analogous to censoring, may lead to an overestimation of the SE obtained from the kernel method. For this reason, we also present two additional indicators that provide information on the SE: ``Regular'' Standard Error (Reg.SE) of each parameter, which is defined as the average SE over the simulation runs where the parameter is significant at a given level (here, $0.05$); Monte Carlo Standard Error (MCSE), that is, the standard error calculated from the coefficient estimates. Finally, the percentage of iteration runs where a given parameter is statistically significant at level $0.05$ is reported (\% sign.). 

The analysis compares the three models under consideration, namely, the data-generating model (ordered logit), linear regression for rank-transformed variables, and mid-quantile regression with $\tau \in\{0.1,0.3, 0.5, 0.7, 0.9\}$. 
For each iteration, the RGA and AGR indices were evaluated on the test dataset. 
The same analysis was subsequently carried out with the real dataset to compare, based on actual evidence, the relative performance of linear regression for rank-transformed variables and mid-quantile regression. 

The use of both quantitative and qualitative regressors mimics the occurrence of exposure (a numerical variable) and attack vector components (factor variables). We generated 
\begin{equation}
    \mathbf{X}_{(cont)}\sim\mathcal{N}(\mu,\sigma),\quad\mathbf{X}_{(cat)}\sim p(\pi_{1},\pi_{2})
\end{equation}
where $\mathbf{X}_{(cont)}$ is a continuous variable with normal distribution $\mathcal{N}(\mu,\sigma)$ with mean $\mu=0$ and variance $\sigma^2=1$; $p(\pi_{1},\pi_{2})$ is the categorical distribution with three support points associated with probability weights $\pi_{1},\pi_{2},1-\pi_{1}-\pi_{2}>0$. In particular, we chose $\pi_{1}=\pi_{2}=\frac{1}{3}$. Then, the responses $y_{i}$, $i\in\{1,\dots,n\}$, were extracted from a categorical distribution with probability derived from (\ref{eq: ordered logit}), i.e., $p(Y=1|\mathbf{X})=P(Y=1|\mathbf{X})$ and 
\begin{equation} 
p(Y=h|\mathbf{X})=P(Y\leq h|\mathbf{X})-P(Y\leq h-1|\mathbf{X}), \quad h\in\{2,\dots,k\}.
\end{equation}
Multiple simulation runs were performed at different choices of $\beta_{\mathrm{true}}$ with different quantile levels.

\subsubsection{Simulation results} 


We start presenting the results of simulations where the response variable contains $k=4$ possible levels. As mentioned above, this situation is in line with the real dataset structure since Tenable's priority rating involves $k=4$ levels too. 

Tables \ref{tab: estimate, k=4 uniform}-\ref{tab: estimates, k=4 non-uniform} report the outcomes from two different scenarios. The parameters defining the theoretical distribution from the OrdLog model can be tuned to obtain the uniform probability distribution on the $k$ response levels (Table \ref{tab: estimate, k=4 uniform}) or they can be chosen generically; in the latter case, we can get a non-uniform distribution (Table \ref{tab: estimates, k=4 non-uniform}). In the tables, we report the estimates of the model parameters (\textrm{Est}) and the corresponding standard errors (\textrm{SE}) averaged over $100$ simulations. We also report the Monte Carlo standard error (\textrm{MCSE}) to evaluate the stability of the estimates over the simulations. For LinReg and MidQR, we report the percentage of times the parameters were significant at the 5\% level (\textrm{\% sign.}). 



\begin{table}[H]
  \centering
  \caption{Coefficient estimates from simulations with $k=4$ levels for the response variable. The parameters in the generative model are tuned in order to get the uniform probability distribution on the $k$ possible response levels.}
  \resizebox{0.9\textwidth}{!}{%
    \begin{tabular}{llccccccc}
\cmidrule{3-9} \morecmidrules \cmidrule{3-9}          &       & \multicolumn{1}{c}{\multirow{2}[3]{*}{$\mathbf{X_{3}}$}} & \multicolumn{1}{c}{\multirow{2}[3]{*}{$\mathbf{X_{4}}$}} & \multicolumn{2}{c}{$\mathbf{X_{1}}$} & \multicolumn{2}{c}{$\mathbf{X_{2}}$} & \multicolumn{1}{c}{\multirow{2}[3]{*}{\textbf{Intercept}}} \\
\cmidrule(r){5-6} \cmidrule(r){7-8}          &       &       &       & \multicolumn{1}{c}{\textbf{1}} & \multicolumn{1}{c}{\textbf{2}} & \multicolumn{1}{c}{\textbf{1}} & \multicolumn{1}{c}{\textbf{2}} &  \\
\cmidrule(r){3-9}
    \multirow{3}[1]{*}{OrdReg} & Est  & -3.097 & 2.094 & 1.017 & 4.141 & -2.062 & 4.227 &  \\
          & SE    & 0.312 & 0.244 & 0.368 & 0.530 & 0.402 & 0.540 &  \\
          & MCSE  & 0.032 & 0.029 & 0.033 & 0.052 & 0.042 & 0.050 &  \\
    \midrule
    \multirow{4}[2]{*}{LinReg} & Est  & -37.012 & 24.156 & 14.856 & 44.762 & -27.017 & 46.337 & 98.235 \\
          & SE    & 2.947 & 2.818 & 7.173 & 7.107 & 7.280 & 7.302 & 6.823 \\
          & MCSE  & 0.230 & 0.235 & 0.566 & 0.626 & 0.651 & 0.544 & 0.489 \\
          & \% sign. & 100.0\% & 100.0\% & 55.0\% & 100.0\% & 99.0\% & 100.0\% & 100.0\% \\
    \midrule
    \multirow{5}[2]{*}{MidQR($\tau_{1}$)} & Est  & -0.238 & 0.156 & 0.038 & 0.359 & -0.146 & 0.482 & 0.291 \\
          & SE    & 2.896 & 2.466 & 7.227 & 6.083 & 7.972 & 6.670 & 7.338 \\
          & Reg.SE & 0.036 & 0.035 & N.D. & 0.086 & 0.092 & 0.090 & 0.089 \\
          & MCSE  & 0.002 & 0.002 & 0.004 & 0.007 & 0.007 & 0.007 & 0.007 \\
          & \% sign. & 71.0\% & 71.0\% & 0.0\% & 70.0\% & 19.0\% & 71.0\% & 66.0\% \\
    \midrule
    \multirow{5}[2]{*}{MidQR($\tau_{2}$)} & Est  & -0.274 & 0.168 & 0.058 & 0.359 & -0.184 & 0.433 & 0.563 \\
          & SE    & 1.283 & 1.192 & 3.365 & 2.648 & 3.563 & 3.178 & 3.150 \\
          & Reg.SE & 0.025 & 0.024 & 0.061 & 0.060 & 0.066 & 0.062 & 0.061 \\
          & MCSE  & 0.002 & 0.002 & 0.005 & 0.006 & 0.008 & 0.006 & 0.006 \\
          & \% sign. & 71.0\% & 71.0\% & 12.0\% & 71.0\% & 57.0\% & 71.0\% & 71.0\% \\
    \midrule
    \multirow{5}[2]{*}{MidQR($\tau_{3}$)} & Est  & -0.270 & 0.163 & 0.046 & 0.300 & -0.188 & 0.344 & 0.827 \\
          & SE    & 705.709 & 340.703 & 372.360 & 1024.919 & 578.466 & 1078.914 & 520.001 \\
          & Reg.SE & 0.022 & 0.021 & 0.058 & 0.056 & 0.061 & 0.056 & 0.057 \\
          & MCSE  & 0.002 & 0.002 & 0.004 & 0.005 & 0.007 & 0.005 & 0.006 \\
          & \% sign. & 54.0\% & 54.0\% & 7.0\% & 54.0\% & 48.0\% & 54.0\% & 54.0\% \\
    \midrule
    \multirow{5}[2]{*}{MidQR($\tau_{4}$)} & Est  & -0.202 & 0.117 & 0.029 & 0.193 & -0.144 & 0.213 & 1.057 \\
          & SE    & 1.267 & 1.148 & 2.258 & 3.299 & 2.433 & 3.350 & 2.410 \\
          & Reg.SE & 0.029 & 0.027 & N.D. & 0.067 & 0.077 & 0.067 & 0.074 \\
          & MCSE  & 0.001 & 0.002 & 0.003 & 0.004 & 0.006 & 0.004 & 0.005 \\
          & \% sign. & 71.0\% & 70.0\% & 0.0\% & 66.0\% & 30.0\% & 67.0\% & 71.0\% \\
    \midrule
    \multirow{5}[2]{*}{MidQR($\tau_{5}$)} & Est  & -0.125 & 0.075 & 0.001 & 0.086 & -0.097 & 0.085 & 1.262 \\
          & SE    & 3.237 & 2.428 & 5.298 & 7.221 & 5.278 & 8.288 & 6.373 \\
          & Reg.SE & 0.040 & 0.034 & N.D. & 0.077 & 0.094 & 0.073 & 0.100 \\
          & MCSE  & 0.001 & 0.001 & 0.002 & 0.003 & 0.004 & 0.003 & 0.004 \\
          & \% sign. & 68.0\% & 46.0\% & 0.0\% & 2.0\% & 1.0\% & 1.0\% & 71.0\% \\
    \Xhline{3.6\arrayrulewidth}
    \end{tabular}
    }
  \label{tab: estimate, k=4 uniform}
\end{table}

\begin{table}[H]
  \centering
  \caption{Coefficient estimates from simulations with $k=4$ levels for the response variable. Generic parameters in the generative model lead to a non-uniform probability distribution on the $k$ possible response levels.}
  \resizebox{0.9\textwidth}{!}{%
    \begin{tabular}{llccccccc}
\cmidrule{3-9} \morecmidrules \cmidrule{3-9}          &       & \multicolumn{1}{c}{\multirow{2}[3]{*}{$\mathbf{X_{3}}$}} & \multicolumn{1}{c}{\multirow{2}[3]{*}{$\mathbf{X_{4}}$}} & \multicolumn{2}{c}{$\mathbf{X_{1}}$} & \multicolumn{2}{c}{$\mathbf{X_{2}}$} & \multicolumn{1}{c}{\multirow{2}[3]{*}{\textbf{Intercept}}} \\
\cmidrule(r){5-6} \cmidrule(r){7-8}          &       &       &       & \multicolumn{1}{c}{\textbf{1}} & \multicolumn{1}{c}{\textbf{2}} & \multicolumn{1}{c}{\textbf{1}} & \multicolumn{1}{c}{\textbf{2}} &  \\
\cmidrule(r){3-9} 
    \multirow{3}[1]{*}{OrdReg} & Est  & -3.116 & 2.064 & 1.046 & 4.120 & -2.074 & 4.094 &  \\
          & SE    & 0.237 & 0.179 & 0.306 & 0.407 & 0.335 & 0.394 &  \\
          & MCSE  & 0.024 & 0.015 & 0.029 & 0.037 & 0.035 & 0.040 &  \\
    \midrule
    \multirow{4}[2]{*}{LinReg} & Est  & -46.974 & 28.359 & 18.304 & 59.905 & -34.439 & 54.792 & 102.372 \\
          & SE    & 2.901 & 2.884 & 6.938 & 7.136 & 7.185 & 7.099 & 6.381 \\
          & MCSE  & 0.269 & 0.225 & 0.670 & 0.612 & 0.645 & 0.609 & 0.506 \\
          & \% sign. & 100.0\% & 100.0\% & 78.0\% & 100.0\% & 100.0\% & 100.0\% & 100.0\% \\
    \midrule
    \multirow{5}[2]{*}{MidQR($\tau_{1}$)} & Est  & -0.311 & 0.166 & 0.083 & 0.385 & -0.140 & 0.453 & 0.288 \\
          & SE    & 3.032 & 2.475 & 6.167 & 6.780 & 7.080 & 6.875 & 7.375 \\
          & Reg.SE & 0.036 & 0.034 & 0.079 & 0.086 & 0.086 & 0.088 & 0.084 \\
          & MCSE  & 0.002 & 0.002 & 0.004 & 0.006 & 0.006 & 0.006 & 0.006 \\
          & \% sign. & 72.0\% & 72.0\% & 2.0\% & 72.0\% & 18.0\% & 72.0\% & 66.0\% \\
    \midrule
    \multirow{5}[2]{*}{MidQR($\tau_{2}$)} & Est  & -0.316 & 0.178 & 0.064 & 0.392 & -0.172 & 0.440 & 0.552 \\
          & SE    & 1.214 & 1.111 & 2.664 & 2.707 & 2.776 & 2.612 & 2.566 \\
          & Reg.SE & 0.023 & 0.023 & 0.057 & 0.057 & 0.061 & 0.058 & 0.056 \\
          & MCSE  & 0.002 & 0.002 & 0.004 & 0.005 & 0.006 & 0.006 & 0.005 \\
          & \% sign. & 72.0\% & 72.0\% & 13.0\% & 72.0\% & 57.0\% & 72.0\% & 72.0\% \\
    \midrule
    \multirow{5}[2]{*}{MidQR($\tau_{3}$)} & Est  & -0.285 & 0.161 & 0.055 & 0.347 & -0.188 & 0.372 & 0.797 \\
          & SE    & 1.303 & 1.756 & 2.397 & 2.540 & 2.926 & 2.497 & 3.089 \\
          & Reg.SE & 0.021 & 0.021 & 0.052 & 0.053 & 0.057 & 0.052 & 0.053 \\
          & MCSE  & 0.002 & 0.002 & 0.004 & 0.004 & 0.006 & 0.005 & 0.005 \\
          & \% sign. & 72.0\% & 72.0\% & 7.0\% & 72.0\% & 68.0\% & 72.0\% & 72.0\% \\
    \midrule
    \multirow{5}[2]{*}{MidQR($\tau_{4}$)} & Est  & -0.202 & 0.114 & 0.038 & 0.244 & -0.147 & 0.249 & 1.023 \\
          & SE    & 1.413 & 1.321 & 2.591 & 3.351 & 2.722 & 3.631 & 2.508 \\
          & Reg.SE & 0.027 & 0.026 & 0.065 & 0.065 & 0.073 & 0.062 & 0.067 \\
          & MCSE  & 0.001 & 0.001 & 0.003 & 0.003 & 0.005 & 0.004 & 0.004 \\
          & \% sign. & 72.0\% & 72.0\% & 1.0\% & 72.0\% & 43.0\% & 71.0\% & 72.0\% \\
    \midrule
    \multirow{5}[2]{*}{MidQR($\tau_{5}$)} & Est  & -0.113 & 0.062 & 0.022 & 0.132 & -0.114 & 0.115 & 1.231 \\
          & SE    & 2.867 & 2.164 & 3.590 & 5.085 & 4.094 & 7.379 & 4.220 \\
          & Reg.SE & 0.038 & 0.034 & N.D. & 0.078 & 0.094 & 0.073 & 0.090 \\
          & MCSE  & 0.002 & 0.001 & 0.002 & 0.003 & 0.004 & 0.003 & 0.004 \\
          & \% sign. & 68.0\% & 25.0\% & 0.0\% & 9.0\% & 2.0\% & 8.0\% & 72.0\% \\
    \Xhline{3.6\arrayrulewidth}
    \end{tabular}
    }
  \label{tab: estimates, k=4 non-uniform}
\end{table}

The resulting RGA and AGR indices are reported in Table \ref{tab: RGA-AGR, k=4}. To provide an informative view of RGA and AGR, we present the boxplots associated with each model in Figure \ref{fig: boxplots RGA-AGR simulations, k=4}. Along with the summary of outputs for the three methods under investigation, in the following tables and figures, we include $\mathrm{RGA}(r_{\mathrm{true}},r_{\mathrm{true}})$ and $\mathrm{AGR}(r_{\mathrm{true}},r_{\mathrm{true}})$ as reference values in the analysis of the two accuracy measures. 

\begin{table}[H]
  \centering
  \caption{RGA and AGR from simulations with $k=4$ levels in the response variable. Columns 2--5 are generated from a model tuned to produce uniform probabilities for the $k$ levels in the response. The last row corresponds to the reference value, namely, the index RGA or AGR evaluated at $(r_{\mathrm{true}},r_{\mathrm{true}})$.}
    \begin{tabular}{crrrrrrrr}
\cmidrule{2-9} \morecmidrules 
\cmidrule{2-9}    \multirow{3}[5]{*}{} & \multicolumn{4}{c}{$k=4$, uniform} & \multicolumn{4}{c}{$k=4$, non-uniform} \\
\cmidrule(r){2-5} \cmidrule(r){6-9}          & \multicolumn{2}{c}{\textbf{RGA}} & \multicolumn{2}{c}{\textbf{AGR}} & \multicolumn{2}{c}{\textbf{RGA}} & \multicolumn{2}{c}{\textbf{AGR}} \\
\cmidrule(r){2-3} \cmidrule(r){4-5} \cmidrule(r){6-7} \cmidrule(r){8-9}         & \multicolumn{1}{c}{Est} & \multicolumn{1}{c}{SD} & \multicolumn{1}{c}{Est} & \multicolumn{1}{c}{SD} & \multicolumn{1}{c}{Est} & \multicolumn{1}{c}{SD} & \multicolumn{1}{c}{Est} & \multicolumn{1}{c}{SD} \\
\hline
    \textrm{OrdLog} & 2.517 & 0.496 & 2.823 & 0.507 & 5.889 & 0.897 & 6.494 & 0.723 \\
    \textrm{LinReg} & 3.276 & 0.578 & 1.516 & 0.193 & 6.762 & 0.796 & 3.254 & 0.282 \\
    \textrm{MidQR}($\tau_{1}$) & 3.093 & 0.551 & 3.016 & 0.394 & 6.600 & 0.767 & 4.316 & 0.348 \\
    \textrm{MidQR}($\tau_{2}$) & 3.212 & 0.555 & 3.143 & 0.391 & 6.657 & 0.768 & 4.356 & 0.342 \\
    \textrm{MidQR}($\tau_{3}$) & 3.239 & 0.562 & 3.214 & 0.389 & 6.684 & 0.773 & 4.377 & 0.343 \\
    \textrm{MidQR}($\tau_{4}$) & 3.193 & 0.565 & 3.207 & 0.398 & 6.670 & 0.797 & 4.371 & 0.349 \\
    \textrm{MidQR}($\tau_{5}$) & 3.016 & 0.573 & 3.146 & 0.418 & 6.491 & 0.862 & 4.276 & 0.370 \\
    $(r_{\mathrm{true}},r_{\mathrm{true}})$  & 4.299 & 0.614 & 4.299 & 0.614 & 8.614 & 0.677 & 8.614 & 0.677 \\
    \Xhline{3.6\arrayrulewidth}
    \end{tabular}
  \label{tab: RGA-AGR, k=4}
\end{table}
\begin{figure}[H]
\begin{subfigure}{.24\textwidth}
  \centering
  \includegraphics[width=\linewidth]{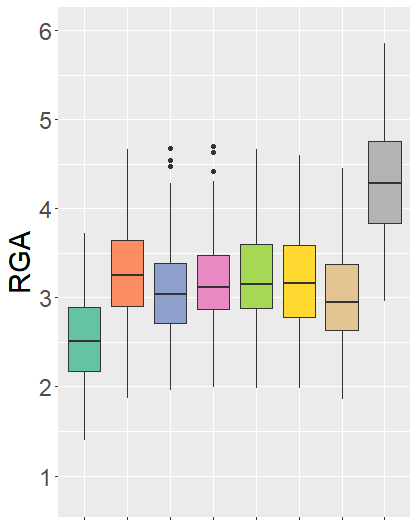}
  \caption{RGA, $k=4$, uniform distribution.}
  \label{fig: boxplots RGA simulations, k=4 uniform}
\end{subfigure}%
\begin{subfigure}{.24\textwidth}
  \centering
  \includegraphics[width=\linewidth]{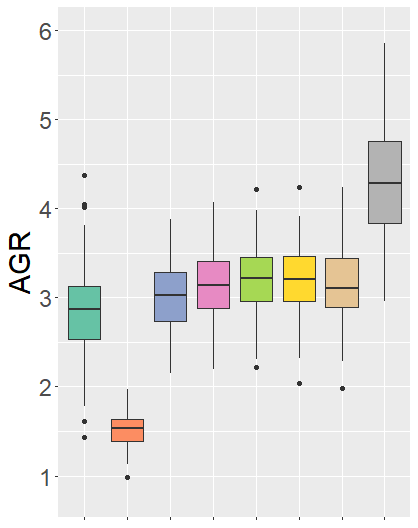}
  \caption{AGR, $k=4$, uniform distribution.}
  \label{fig: boxplots AGR simulations, k=4 uniform}
\end{subfigure}
\begin{subfigure}{.24\textwidth}
  \centering
  \includegraphics[width=\linewidth]{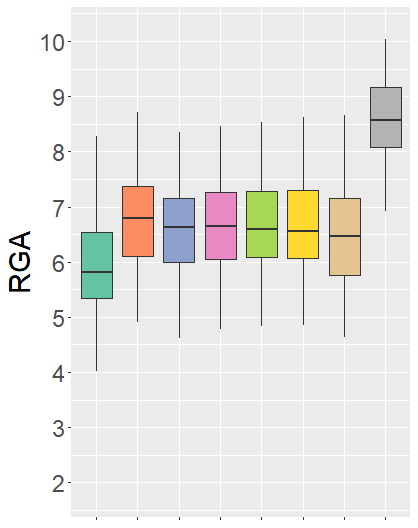}
  \caption{RGA, $k=4$, non-uniform distribution.}
  \label{fig: boxplots RGA simulations, k=4 non-uniform}
\end{subfigure}
\begin{subfigure}{.24\textwidth}
  \centering
  \includegraphics[width=\linewidth]{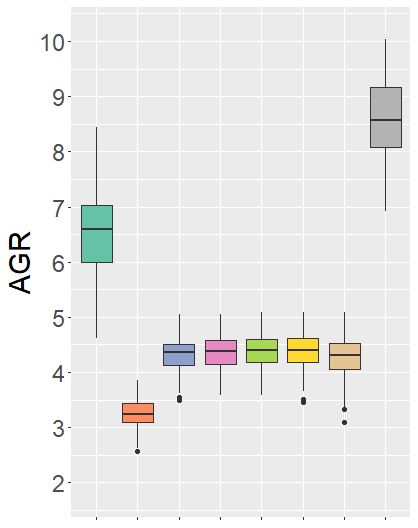}
  \caption{AGR, $k=4$, non-uniform distribution.}
  \label{fig: boxplots AGR simulations, k=4 non-uniform}
\end{subfigure}
\caption{Boxplots for RGA and AGR when $k=4$; both uniform and non-uniform probability distributions are considered starting from the data-generating OrdLog model. Boxplots refer, from left to right of the x-axis, to OrdLog, LinReg, MidQR with $\tau$ taking values in $\{0.1, 0.3, 0.5, 0.7, 0.9\}$, and the reference value $\mathrm{RGA}(r_{\mathrm{true}},r_{\mathrm{true}})$.}
\label{fig: boxplots RGA-AGR simulations, k=4}
\end{figure} 


Then, we move to different numbers of levels in order to better assess the behaviour of the different methods in different decision scenarios. We address this aspect starting with $k=3$: this is a typical scale in several operational or tactical decisions, where levels are generally interpreted as ``low'', ``medium'', and ``high'', respectively. The outcomes of this set of simulations are presented in Table \ref{tab: estimates, k=3 non-uniform}. 
\begin{table}[H]
  \centering
  \caption{Coefficient estimates from simulations with $k=3$ levels for the response variable.}
  \resizebox{0.9\textwidth}{!}{%
    \begin{tabular}{llccccccc}
\cmidrule{3-9} \morecmidrules \cmidrule{3-9}          &       & \multicolumn{1}{c}{\multirow{2}[3]{*}{$\mathbf{X_{3}}$}} & \multicolumn{1}{c}{\multirow{2}[3]{*}{$\mathbf{X_{4}}$}} & \multicolumn{2}{c}{$\mathbf{X_{1}}$} & \multicolumn{2}{c}{$\mathbf{X_{2}}$} & \multicolumn{1}{c}{\multirow{2}[3]{*}{\textbf{Intercept}}} \\
\cmidrule(r){5-6} \cmidrule(r){7-8}          &       &       &       & \multicolumn{1}{c}{\textbf{1}} & \multicolumn{1}{c}{\textbf{2}} & \multicolumn{1}{c}{\textbf{1}} & \multicolumn{1}{c}{\textbf{2}} &  \\
\cmidrule(r){3-9}
    \multirow{3}[1]{*}{OrdReg} & Est  & -3.173 & 2.083 & 1.053 & 4.249 & -2.086 & 4.193 &  \\
          & SE    & 0.395 & 0.298 & 0.466 & 0.745 & 0.499 & 0.755 &  \\
          & MCSE  & 0.038 & 0.028 & 0.050 & 0.072 & 0.042 & 0.082 &  \\
    \midrule
    \multirow{4}[2]{*}{LinReg} & Est  & -23.122 & 15.755 & 9.877 & 28.192 & -17.554 & 24.575 & 74.764 \\
          & SE    & 1.825 & 1.827 & 4.439 & 4.732 & 4.609 & 4.568 & 4.152 \\
          & MCSE  & 0.199 & 0.168 & 0.379 & 0.403 & 0.395 & 0.346 & 0.418 \\
          & \% sign. & 100.0\% & 100.0\% & 69.0\% & 100.0\% & 99.0\% & 100.0\% & 100.0\% \\
    \midrule
    \multirow{5}[2]{*}{MidQR($\tau_{1}$)} & Est  & -0.195 & 0.114 & 0.038 & 0.270 & -0.129 & 0.291 & 0.341 \\
          & SE    & 12.519 & 16.381 & 27.625 & 34.324 & 37.952 & 25.530 & 31.305 \\
          & Reg.SE & 0.027 & 0.028 & N.D. & 0.072 & 0.072 & 0.071 & 0.070 \\
          & MCSE  & 0.002 & 0.002 & 0.003 & 0.004 & 0.004 & 0.005 & 0.004 \\
          & \% sign. & 70.0\% & 69.0\% & 0.0\% & 70.0\% & 25.0\% & 70.0\% & 70.0\% \\
    \midrule
    \multirow{5}[2]{*}{MidQR($\tau_{2}$)} & Est  & -0.218 & 0.138 & 0.047 & 0.259 & -0.133 & 0.277 & 0.550 \\
          & SE    & 8.409 & 6.741 & 17.770 & 18.895 & 19.943 & 18.942 & 16.774 \\
          & Reg.SE & 0.019 & 0.019 & 0.049 & 0.050 & 0.054 & 0.048 & 0.049 \\
          & MCSE  & 0.001 & 0.002 & 0.003 & 0.004 & 0.005 & 0.004 & 0.004 \\
          & \% sign. & 70.0\% & 70.0\% & 8.0\% & 70.0\% & 50.0\% & 70.0\% & 70.0\% \\
    \midrule
    \multirow{5}[2]{*}{MidQR($\tau_{3}$)} & Est  & -0.206 & 0.134 & 0.056 & 0.219 & -0.133 & 0.222 & 0.765 \\
          & SE    & 753.120 & 344.070 & 171.833 & 819.444 & 253.610 & 970.775 & 573.024 \\
          & Reg.SE & 0.019 & 0.018 & 0.046 & 0.046 & 0.050 & 0.042 & 0.045 \\
          & MCSE  & 0.002 & 0.002 & 0.003 & 0.004 & 0.005 & 0.004 & 0.004 \\
          & \% sign. & 60.0\% & 60.0\% & 16.0\% & 60.0\% & 49.0\% & 60.0\% & 60.0\% \\
    \midrule
    \multirow{5}[2]{*}{MidQR($\tau_{4}$)} & Est  & -0.129 & 0.087 & 0.040 & 0.121 & -0.086 & 0.116 & 0.924 \\
          & SE    & 22.573 & 11.780 & 28.125 & 42.421 & 27.902 & 57.145 & 31.146 \\
          & Reg.SE & 0.028 & 0.024 & N.D. & 0.058 & 0.061 & 0.053 & 0.061 \\
          & MCSE  & 0.001 & 0.001 & 0.002 & 0.003 & 0.004 & 0.003 & 0.003 \\
          & \% sign. & 69.0\% & 67.0\% & 0.0\% & 43.0\% & 13.0\% & 25.0\% & 70.0\% \\
    \midrule
    \multirow{5}[2]{*}{MidQR($\tau_{5}$)} & Est  & -0.061 & 0.042 & 0.029 & 0.045 & -0.045 & 0.036 & 1.036 \\
          & SE    & 48.199 & 25.136 & 34.366 & 61.267 & 41.685 & 82.775 & 74.481 \\
          & Reg.SE & 0.030 & 0.027 & N.D. & 0.060 & N.D. & N.D. & 0.119 \\
          & MCSE  & 0.001 & 0.001 & 0.002 & 0.002 & 0.003 & 0.001 & 0.002 \\
          & \% sign. & 20.0\% & 10.0\% & 0.0\% & 1.0\% & 0.0\% & 0.0\% & 70.0\% \\
    \Xhline{3.6\arrayrulewidth}
    \end{tabular}
    }
  \label{tab: estimates, k=3 non-uniform}
\end{table}
The corresponding RGA and AGR indices are shown in Table \ref{tab: RGA-AGR, k=3}. Even in this case, we provide a graphical representation of these outcomes in Figure \ref{fig: boxplots RGA-AGR simulations, k=3}.
\begin{table}[H]
\begin{minipage}{.5\linewidth}
  \centering
  \captionof{table}{RGA and AGR from simulations with a low number $k=3$ of levels for the response variable. The last row corresponds to the reference value, namely, the index RGA or AGR evaluated at $(r_{\mathrm{true}},r_{\mathrm{true}})$.}
    \begin{tabular}{lcccc}
\cmidrule{2-5} \morecmidrules \cmidrule{2-5}   \multirow{2}[4]{*}{} & \multicolumn{2}{c}{\textbf{RGA}} & \multicolumn{2}{c}{\textbf{AGR}} \\
\cmidrule(r){2-3} \cmidrule(r){4-5}          & \multicolumn{1}{c}{Est} & \multicolumn{1}{c}{SD} & \multicolumn{1}{c}{Est} & \multicolumn{1}{c}{SD} \\
    \midrule
    \textrm{OrdLog} & 1.439 & 0.488 & 1.545 & 0.538 \\
    \textrm{LinReg} & 2.203 & 0.667 & 0.865 & 0.169 \\
    \textrm{MidQR}($\tau_{1}$) & 2.113 & 0.677 & 2.733 & 0.487 \\
    \textrm{MidQR}($\tau_{2}$) & 2.193 & 0.677 & 2.871 & 0.473 \\
    \textrm{MidQR}($\tau_{3}$) & 2.162 & 0.677 & 2.883 & 0.470 \\
    \textrm{MidQR}($\tau_{4}$) & 2.082 & 0.667 & 2.848 & 0.470 \\
    \textrm{MidQR}($\tau_{5}$) & 1.785 & 0.616 & 2.631 & 0.468 \\
    $(r_{\mathrm{true}},r_{\mathrm{true}})$  & 3.499 & 0.819 & 3.499 & 0.819 \\
    \Xhline{3.6\arrayrulewidth}
    \end{tabular}
  \label{tab: RGA-AGR, k=3}
  \end{minipage}
\begin{minipage}{.48\linewidth}
\begin{subtable}{.5\textwidth}
  \centering
  \includegraphics[width=\linewidth]{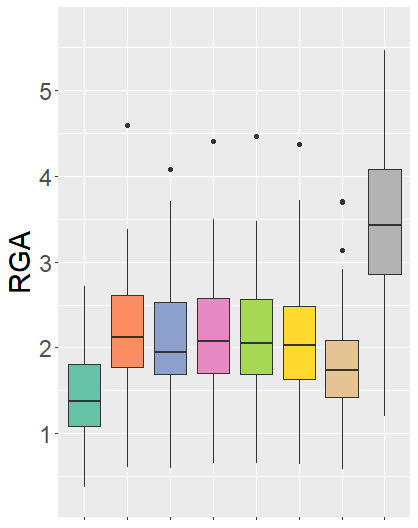}
  \caption{RGA, $k=3$.}
  \label{fig: boxplots RGA simulations, k=3}
\end{subtable}%
\hfill
\begin{subtable}{.5\textwidth}
  \centering
  \includegraphics[width=\linewidth]{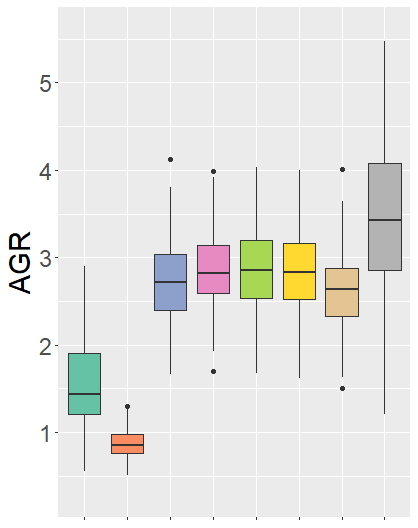}
  \caption{AGR, $k=3$.}
  \label{fig: boxplots AGR simulations, k=3}
\end{subtable}
\captionof{figure}{Boxplots for RGA and AGR when $k=3$. Boxplots refer, from left to right of the x-axis, to OrdLog, LinReg, MidQR with $\tau\in\{0.1, 0.3, 0.5, 0.7, 0.9\}$, and 
the reference value 
$\mathrm{RGA}(r_{\mathrm{true}},r_{\mathrm{true}})$.}
\label{fig: boxplots RGA-AGR simulations, k=3}
\end{minipage}
\end{table}

Finally, we complete the simulation study by considering more than $4$ levels in the response variable. Specifically, we report the results at $k=6$ (Table \ref{tab: estimates, k=6 non-uniform}) and $k=8$ (Table \ref{tab: estimates, k=8 non-uniform}). The boxplots corresponding to the RGA and AGR indices summarised in Table \ref{tab: RGA-AGR, k=8 and k=9} are displayed in Figure \ref{fig: boxplots RGA-AGR simulations, k=6 and k=8}. 

\begin{figure}[H]
\begin{subfigure}{.24\textwidth}
  \centering
  \includegraphics[width=\linewidth]{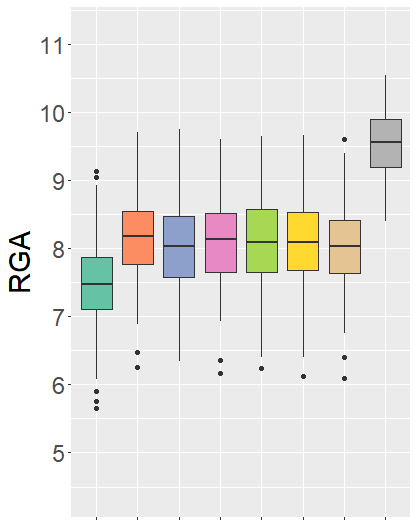}
  \caption{RGA, $k=6$.}
  \label{fig: boxplots RGA simulations, k=6}
\end{subfigure}
\begin{subfigure}{.24\textwidth}
  \centering
  \includegraphics[width=\linewidth]{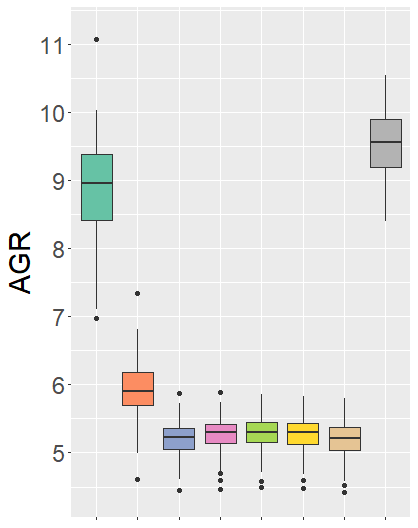}
  \caption{AGR, $k=6$.}
  \label{fig: boxplots AGR simulations, k=6}
\end{subfigure}
\begin{subfigure}{.24\textwidth}
  \centering
  \includegraphics[width=\linewidth]{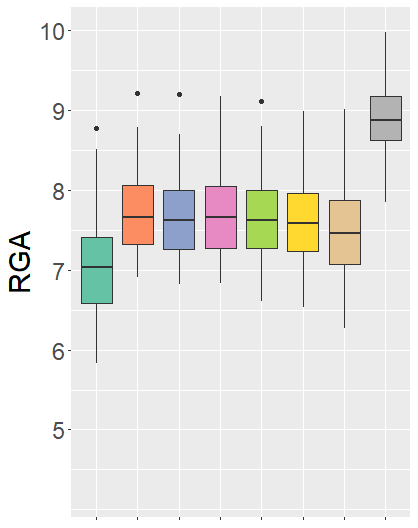}
  \caption{RGA, $k=8$.}
  \label{fig: boxplots RGA simulations, k=8 non-uniform}
\end{subfigure}
\begin{subfigure}{0.24\textwidth}
  \centering
  \includegraphics[width=\linewidth]{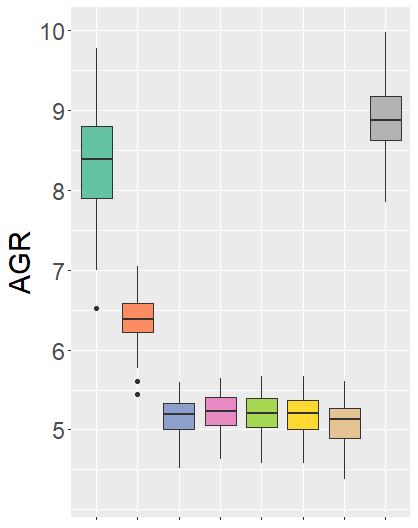}
  \caption{AGR, $k=8$.}
  \label{fig: boxplots AGR simulations, k=8}
\end{subfigure}
\caption{Boxplots for RGA and AGR when $k=6$ or $k=8$. Boxplots refer, from left to right of the x-axis, to OrdLog, LinReg, MidQR with $\tau$ taking values in $\{0.1, 0.3, 0.5, 0.7, 0.9\}$, and the reference value $\mathrm{RGA}(r_{\mathrm{true}},r_{\mathrm{true}})$.}
\label{fig: boxplots RGA-AGR simulations, k=6 and k=8}
\end{figure} 

\begin{table}[H]
  \centering
  \caption{Coefficient estimates from simulations with $k=6$ levels for the response variable.}
  \resizebox{0.9\textwidth}{!}{%
    \begin{tabular}{llccccccc}
\cmidrule{3-9} \morecmidrules \cmidrule{3-9}          &       & \multicolumn{1}{c}{\multirow{2}[3]{*}{$\mathbf{X_{3}}$}} & \multicolumn{1}{c}{\multirow{2}[3]{*}{$\mathbf{X_{4}}$}} & \multicolumn{2}{c}{$\mathbf{X_{1}}$} & \multicolumn{2}{c}{$\mathbf{X_{2}}$} & \multicolumn{1}{c}{\multirow{2}[3]{*}{\textbf{Intercept}}} \\
\cmidrule(r){5-6} \cmidrule(r){7-8}          &       &       &       & \multicolumn{1}{c}{\textbf{1}} & \multicolumn{1}{c}{\textbf{2}} & \multicolumn{1}{c}{\textbf{1}} & \multicolumn{1}{c}{\textbf{2}} &  \\
\cmidrule(r){3-9}    \multirow{3}[1]{*}{OrdReg} & Est  & -3.116 & 2.064 & 1.046 & 4.120 & -2.074 & 4.094 &  \\
          & SE    & 0.237 & 0.179 & 0.306 & 0.407 & 0.335 & 0.394 &  \\
          & MCSE  & 0.024 & 0.015 & 0.029 & 0.037 & 0.035 & 0.040 &  \\
    \midrule
    \multirow{4}[2]{*}{LinReg} & Est  & -61.725 & 41.627 & 23.455 & 76.997 & -48.392 & 80.101 & 108.517 \\
          & SE    & 3.038 & 2.943 & 7.577 & 7.588 & 7.754 & 7.355 & 6.525 \\
          & MCSE  & 0.202 & 0.230 & 0.635 & 0.603 & 0.716 & 0.624 & 0.521 \\
          & \% sign. & 100.0\% & 100.0\% & 89.0\% & 100.0\% & 100.0\% & 100.0\% & 100.0\% \\
    \midrule
    \multirow{5}[2]{*}{MidQR($\tau_{1}$)} & Est  & -0.347 & 0.217 & 0.007 & 0.387 & -0.203 & 0.532 & 0.366 \\
          & SE    & 0.821 & 0.717 & 2.078 & 2.096 & 2.573 & 1.958 & 2.207 \\
          & Reg.SE & 0.033 & 0.033 & N.D. & 0.084 & 0.090 & 0.084 & 0.078 \\
          & MCSE  & 0.001 & 0.002 & 0.004 & 0.006 & 0.005 & 0.006 & 0.004 \\
          & \% sign. & 89.0\% & 89.0\% & 0.0\% & 89.0\% & 58.0\% & 89.0\% & 89.0\% \\
    \midrule
    \multirow{5}[2]{*}{MidQR($\tau_{2}$)} & Est  & -0.342 & 0.230 & 0.075 & 0.404 & -0.254 & 0.518 & 0.597 \\
          & SE    & 0.388 & 0.395 & 0.984 & 0.993 & 1.054 & 0.944 & 0.935 \\
          & Reg.SE & 0.023 & 0.022 & 0.056 & 0.058 & 0.063 & 0.055 & 0.051 \\
          & MCSE  & 0.001 & 0.002 & 0.004 & 0.005 & 0.005 & 0.005 & 0.004 \\
          & \% sign. & 89.0\% & 89.0\% & 18.0\% & 89.0\% & 89.0\% & 89.0\% & 89.0\% \\
    \midrule
    \multirow{5}[2]{*}{MidQR($\tau_{3}$)} & Est  & -0.314 & 0.213 & 0.089 & 0.363 & -0.230 & 0.437 & 0.830 \\
          & SE    & 2.967 & 2.491 & 2.748 & 5.410 & 1.992 & 6.140 & 4.738 \\
          & Reg.SE & 0.023 & 0.021 & 0.053 & 0.053 & 0.062 & 0.049 & 0.052 \\
          & MCSE  & 0.002 & 0.002 & 0.004 & 0.005 & 0.005 & 0.004 & 0.004 \\
          & \% sign. & 81.0\% & 80.0\% & 31.0\% & 81.0\% & 83.0\% & 80.0\% & 85.0\% \\
    \midrule
    \multirow{5}[2]{*}{MidQR($\tau_{4}$)} & Est  & -0.253 & 0.173 & 0.080 & 0.285 & -0.192 & 0.337 & 1.077 \\
          & SE    & 0.320 & 0.293 & 0.689 & 0.675 & 0.695 & 0.688 & 0.555 \\
          & Reg.SE & 0.026 & 0.025 & 0.066 & 0.064 & 0.075 & 0.057 & 0.064 \\
          & MCSE  & 0.001 & 0.002 & 0.003 & 0.004 & 0.005 & 0.003 & 0.004 \\
          & \% sign. & 89.0\% & 89.0\% & 8.0\% & 89.0\% & 69.0\% & 89.0\% & 89.0\% \\
    \midrule
    \multirow{5}[2]{*}{MidQR($\tau_{5}$)} & Est  & -0.185 & 0.130 & 0.059 & 0.186 & -0.131 & 0.219 & 1.347 \\
          & SE    & 0.500 & 0.409 & 0.962 & 1.050 & 0.936 & 1.179 & 0.846 \\
          & Reg.SE & 0.036 & 0.035 & N.D. & 0.086 & 0.103 & 0.075 & 0.090 \\
          & MCSE  & 0.001 & 0.002 & 0.003 & 0.004 & 0.005 & 0.003 & 0.004 \\
          & \% sign. & 89.0\% & 89.0\% & 0.0\% & 58.0\% & 6.0\% & 88.0\% & 89.0\% \\
    \Xhline{3.6\arrayrulewidth}
    \end{tabular}
    }
  \label{tab: estimates, k=6 non-uniform}
\end{table}

\begin{table}[H]
  \centering
  \caption{Coefficient estimates from simulations with $k=8$ levels for the response variable.}
  \resizebox{0.9\textwidth}{!}{%
    \begin{tabular}{llccccccc}
\cmidrule{3-9} \morecmidrules \cmidrule{3-9}          &       & \multicolumn{1}{c}{\multirow{2}[3]{*}{$\mathbf{X_{3}}$}} & \multicolumn{1}{c}{\multirow{2}[3]{*}{$\mathbf{X_{4}}$}} & \multicolumn{2}{c}{$\mathbf{X_{1}}$} & \multicolumn{2}{c}{$\mathbf{X_{2}}$} & \multicolumn{1}{c}{\multirow{2}[3]{*}{\textbf{Intercept}}} \\
\cmidrule(r){5-6} \cmidrule(r){7-8}          &       &       &       & \multicolumn{1}{c}{\textbf{1}} & \multicolumn{1}{c}{\textbf{2}} & \multicolumn{1}{c}{\textbf{1}} & \multicolumn{1}{c}{\textbf{2}} &  \\
\cmidrule(r){3-9}    \multirow{3}[1]{*}{OrdReg} & Est  & -3.062 & 2.053 & 1.008 & 4.047 & -2.045 & 4.040 &  \\
          & SE    & 0.217 & 0.170 & 0.289 & 0.373 & 0.305 & 0.363 &  \\
          & MCSE  & 0.021 & 0.019 & 0.027 & 0.033 & 0.031 & 0.037 &  \\
    \midrule
    \multirow{4}[2]{*}{LinReg} & Est  & -67.507 & 44.804 & 24.680 & 90.220 & -44.497 & 92.607 & 111.317 \\
          & SE    & 2.987 & 2.963 & 7.007 & 7.248 & 6.947 & 6.999 & 6.752 \\
          & MCSE  & 0.202 & 0.301 & 0.634 & 0.614 & 0.650 & 0.613 & 0.559 \\
          & \% sign. & 100.0\% & 100.0\% & 95.0\% & 100.0\% & 100.0\% & 100.0\% & 100.0\% \\
    \midrule
    \multirow{5}[2]{*}{MidQR($\tau_{1}$)} & Est  & -0.414 & 0.241 & 0.127 & 0.585 & -0.267 & 0.596 & 0.488 \\
          & SE    & 6.789 & 2.250 & 12.100 & 12.866 & 6.430 & 6.874 & 12.916 \\
          & Reg.SE & 0.036 & 0.037 & 0.087 & 0.091 & 0.093 & 0.085 & 0.094 \\
          & MCSE  & 0.002 & 0.002 & 0.004 & 0.007 & 0.008 & 0.007 & 0.007 \\
          & \% sign. & 73.0\% & 73.0\% & 14.0\% & 73.0\% & 61.0\% & 73.0\% & 73.0\% \\
    \midrule
    \multirow{5}[2]{*}{MidQR($\tau_{2}$)} & Est  & -0.409 & 0.266 & 0.095 & 0.537 & -0.271 & 0.525 & 0.811 \\
          & SE    & 1.017 & 0.950 & 2.209 & 2.224 & 2.397 & 2.045 & 2.049 \\
          & Reg.SE & 0.025 & 0.024 & 0.061 & 0.061 & 0.062 & 0.057 & 0.062 \\
          & MCSE  & 0.002 & 0.002 & 0.004 & 0.006 & 0.007 & 0.005 & 0.006 \\
          & \% sign. & 74.0\% & 74.0\% & 23.0\% & 74.0\% & 73.0\% & 74.0\% & 74.0\% \\
    \midrule
    \multirow{5}[2]{*}{MidQR($\tau_{3}$)} & Est  & -0.363 & 0.250 & 0.048 & 0.436 & -0.251 & 0.427 & 1.090 \\
          & SE    & 0.988 & 0.983 & 1.959 & 1.558 & 1.624 & 1.879 & 1.754 \\
          & Reg.SE & 0.024 & 0.024 & 0.061 & 0.058 & 0.060 & 0.051 & 0.062 \\
          & MCSE  & 0.002 & 0.002 & 0.004 & 0.005 & 0.006 & 0.004 & 0.005 \\
          & \% sign. & 74.0\% & 74.0\% & 6.0\% & 74.0\% & 72.0\% & 74.0\% & 74.0\% \\
    \midrule
    \multirow{5}[2]{*}{MidQR($\tau_{4}$)} & Est  & -0.297 & 0.208 & 0.021 & 0.337 & -0.219 & 0.335 & 1.339 \\
          & SE    & 0.729 & 0.652 & 1.655 & 1.725 & 1.645 & 1.672 & 1.582 \\
          & Reg.SE & 0.031 & 0.030 & N.D. & 0.070 & 0.074 & 0.059 & 0.077 \\
          & MCSE  & 0.002 & 0.002 & 0.003 & 0.004 & 0.006 & 0.004 & 0.005 \\
          & \% sign. & 74.0\% & 74.0\% & 0.0\% & 74.0\% & 68.0\% & 73.0\% & 74.0\% \\
    \midrule
    \multirow{5}[2]{*}{MidQR($\tau_{5}$)} & Est  & -0.221 & 0.154 & -0.004 & 0.220 & -0.185 & 0.212 & 1.628 \\
          & SE    & 1.905 & 1.125 & 1.831 & 3.658 & 2.830 & 2.480 & 2.603 \\
          & Reg.SE & 0.042 & 0.041 & N.D. & 0.094 & 0.100 & 0.080 & 0.106 \\
          & MCSE  & 0.002 & 0.002 & 0.002 & 0.003 & 0.005 & 0.003 & 0.004 \\
          & \% sign. & 74.0\% & 74.0\% & 0.0\% & 65.0\% & 23.0\% & 70.0\% & 74.0\% \\
    \Xhline{3.6\arrayrulewidth}
    \end{tabular}
    }
  \label{tab: estimates, k=8 non-uniform}
\end{table}

\begin{table}[H]
  \centering
  \caption{RGA and AGR from simulations with a higher number of levels for the response variable: $k=6$ (columns 2-5) and $k=8$ (columns 6-9). The last row corresponds to the reference value, namely, the index RGA or AGR evaluated at $(r_{\mathrm{true}},r_{\mathrm{true}})$.}
    \begin{tabular}{lcccccccc}
\cmidrule{2-9} \morecmidrules \cmidrule{2-9}
    \multirow{3}[5]{*}{} & \multicolumn{4}{c}{$k=6$}       & \multicolumn{4}{c}{$k=8$} \\
\cmidrule(r){2-5} \cmidrule(r){6-9}          & \multicolumn{2}{c}{\textbf{RGA}} & \multicolumn{2}{c}{\textbf{AGR}} & \multicolumn{2}{c}{\textbf{RGA}} & \multicolumn{2}{c}{\textbf{AGR}} \\
\cmidrule(r){2-3} \cmidrule(r){4-5} \cmidrule(r){6-7} \cmidrule(r){8-9}          & \multicolumn{1}{c}{Est} & \multicolumn{1}{c}{SD} & \multicolumn{1}{c}{Est} & \multicolumn{1}{c}{SD} & \multicolumn{1}{c}{Est} & \multicolumn{1}{c}{SD} & \multicolumn{1}{c}{Est} & \multicolumn{1}{c}{SD} \\
    \midrule
    OrdLog & 7.468 & 0.679 & 8.865 & 0.717 & 6.999 & 0.603 & 8.344 & 0.644 \\
    LinReg & 8.124 & 0.652 & 5.932 & 0.426 & 7.709 & 0.494 & 6.365 & 0.303 \\
    MidQR($\tau_{1}$) & 8.025 & 0.683 & 5.206 & 0.248 & 7.636 & 0.495 & 5.164 & 0.234 \\
    MidQR($\tau_{2}$) & 8.064 & 0.664 & 5.268 & 0.246 & 7.682 & 0.493 & 5.222 & 0.221 \\
    MidQR($\tau_{3}$) & 8.080 & 0.661 & 5.268 & 0.249 & 7.641 & 0.513 & 5.206 & 0.237 \\
    MidQR($\tau_{4}$) & 8.067 & 0.657 & 5.256 & 0.253 & 7.598 & 0.510 & 5.177 & 0.241 \\
    MidQR($\tau_{5}$) & 7.989 & 0.645 & 5.183 & 0.273 & 7.475 & 0.558 & 5.080 & 0.267 \\
    $(r_{\mathrm{true}},r_{\mathrm{true}})$  & 9.533 & 0.515 & 9.533 & 0.515 & 8.932 & 0.436 & 8.932 & 0.436 \\
    \Xhline{3.6\arrayrulewidth}
    \end{tabular}
  \label{tab: RGA-AGR, k=8 and k=9}
\end{table}

\subsubsection{Real dataset analysis}

In parallel with the investigation of the simulated data, we report the study of the dataset whose construction has been described in Section \ref{sec: data sources}. In particular, we present the same type of indicators considered for the simulations. However, here we stress that multiple datasets are constructed from the original one through its random splitting into a training set ($n_{\mathrm{tr}}=664$) and a test set ($n_{\mathrm{test}}=50$). This splitting of the dataset takes into account the imbalance of cyber vulnerability characteristics, so a smaller percentage of observations in the training set could cause the models, in principle, to miss relevant information about rare events. This aspect also occurs in other statistical analyses of cybersecurity \citep{Giudici2021}. 

We generated $100$ random extraction of test sets, whose complements return the associated training sets, to evaluate averaged parameter estimates, standard errors, and predictive performance indices; $16$ quantile levels equally spaced between $0.1$ and $0.9$ are considered in this case.  

We start with parameter estimates, which are shown in Table \ref{tab: estimates, real_all}. Here, the whole set of variables described in Table \ref{tab: data structure} is used to implement the regression models. Then we restrict these models by considering only technical ($X_{\mathrm{AC}}$, $X_{\mathrm{AV}}$) and contextual (exposure, exploit) variables; the corresponding outcomes are presented in Table \ref{tab: estimates, real_Techn}. 

\begin{sidewaystable}[htbp]
  \centering
  \caption{Parameter estimates from data regarding real cyber-vulnerabilities. All the variables have been used as regressors.}
  
\begin{adjustbox}{width=1\textwidth}
    \begin{tabular}{clrrrrrrrrrrrrccc}
\cmidrule{3-17}    \multicolumn{2}{c}{\multirow{2}[4]{*}{}} & \multicolumn{1}{c}{\textbf{Exposure}} & \multicolumn{2}{c}{\textbf{C}} & \multicolumn{2}{c}{\textbf{I}} & \multicolumn{2}{c}{\textbf{A}} & \multicolumn{2}{c}{\textbf{AV}} & \multicolumn{2}{c}{\textbf{AC}} & \multicolumn{1}{c}{\textbf{Exploit}} & \multicolumn{3}{c}{\textbf{Intercept(s)}} \\
\cmidrule{3-17}    \multicolumn{2}{c}{} &       & \multicolumn{1}{c}{\textbf{L}} & \multicolumn{1}{c}{\textbf{Q}} & \multicolumn{1}{c}{\textbf{L}} & \multicolumn{1}{c}{\textbf{Q}} & \multicolumn{1}{c}{\textbf{L}} & \multicolumn{1}{c}{\textbf{Q}} & \multicolumn{1}{c}{\textbf{L}} & \multicolumn{1}{c}{\textbf{Q}} & \multicolumn{1}{c}{\textbf{L}} & \multicolumn{1}{c}{\textbf{Q}} &       & \textbf{1|2} & \textbf{2|3} & \textbf{3|4} \\
        \midrule
    \multirow{3}[2]{*}{OrdReg} & Mean  & -0.002 & -0.292 & -0.360 & 1.034 & 0.268 & 0.588 & -0.236 & 0.014 & 0.589 & -0.094 & 0.127 & 0.201 & \multicolumn{1}{r}{-2.560} & \multicolumn{1}{r}{1.125} & \multicolumn{1}{r}{3.301} \\
          & SE    & 0.009 & 0.840 & 0.506 & 0.893 & 0.540 & 0.437 & 0.272 & 0.471 & 0.281 & 0.286 & 0.211 & 0.218 & \multicolumn{1}{r}{0.432} & \multicolumn{1}{r}{0.420} & \multicolumn{1}{r}{0.472} \\
          & MCSE  & 0.00028 & 0.02789 & 0.01665 & 0.03199 & 0.01955 & 0.01509 & 0.00856 & 0.02032 & 0.01150 & 0.00856 & 0.00574 & 0.00636 & \multicolumn{1}{r}{0.01586} & \multicolumn{1}{r}{0.01595} & \multicolumn{1}{r}{0.01804} \\
    \midrule
    \midrule
    \multirow{3}[2]{*}{LinReg} & Mean  & -1.960 & -103.338 & -54.922 & 141.112 & 59.905 & 18.869 & -4.777 & -44.815 & 90.601 & -14.894 & 18.076 & 16.066 & \multicolumn{3}{c}{305.143} \\
          & SE    & 0.805 & 84.906 & 50.789 & 88.633 & 53.262 & 38.948 & 24.200 & 43.178 & 25.746 & 26.073 & 19.402 & 20.194 & \multicolumn{3}{c}{37.551} \\
          & MCSE  & 0.02322 & 2.51845 & 1.49273 & 2.77951 & 1.68523 & 1.38273 & 0.80690 & 1.66092 & 0.97357 & 0.70261 & 0.52549 & 0.57981 & \multicolumn{3}{c}{1.27969} \\
    \midrule
    \midrule
    \multirow{3}[2]{*}{MidQR($\tau_{1}$)} & Mean  & 0.002 & 0.032 & -0.020 & 0.053 & -0.015 & 0.047 & -0.018 & 0.025 & 0.024 & 0.006 & -0.010 & 0.006 & \multicolumn{3}{c}{0.083} \\
          & SE    & 0.024 & 2.301 & 1.406 & 2.544 & 1.553 & 1.463 & 0.906 & 2.557 & 1.484 & 0.903 & 0.642 & 0.636 & \multicolumn{3}{c}{1.526} \\
          & MCSE  & 0.00002 & 0.00151 & 0.00093 & 0.00200 & 0.00115 & 0.00060 & 0.00035 & 0.00091 & 0.00063 & 0.00042 & 0.00033 & 0.00032 & \multicolumn{3}{c}{0.00091} \\
    \midrule
    \multirow{3}[2]{*}{MidQR($\tau_{4}$)} & Mean 4 & 0.000 & 0.015 & -0.044 & 0.068 & -0.009 & 0.073 & -0.023 & 0.043 & 0.041 & 0.014 & -0.008 & 0.006 & \multicolumn{3}{c}{0.412} \\
          & SE 4  & 0.020 & 2.427 & 1.450 & 2.748 & 1.656 & 1.236 & 0.757 & 1.901 & 1.100 & 0.794 & 0.593 & 0.602 & \multicolumn{3}{c}{1.181} \\
          & MCSE 4 & 0.00003 & 0.00241 & 0.00154 & 0.00286 & 0.00175 & 0.00101 & 0.00044 & 0.00151 & 0.00085 & 0.00054 & 0.00044 & 0.00050 & \multicolumn{3}{c}{0.00117} \\
    \midrule
    \multirow{3}[2]{*}{MidQR($\tau_{7}$)} & Mean 7 & -0.001 & 0.000 & -0.040 & 0.067 & -0.003 & 0.059 & -0.016 & 0.031 & 0.043 & 0.018 & -0.007 & 0.006 & \multicolumn{3}{c}{0.672} \\
          & SE 7  & 0.017 & 1.795 & 1.092 & 1.951 & 1.189 & 0.982 & 0.607 & 1.458 & 0.845 & 0.663 & 0.496 & 0.489 & \multicolumn{3}{c}{0.940} \\
          & MCSE 7 & 0.00003 & 0.00191 & 0.00128 & 0.00220 & 0.00140 & 0.00094 & 0.00040 & 0.00131 & 0.00072 & 0.00044 & 0.00039 & 0.00046 & \multicolumn{3}{c}{0.00110} \\
    \midrule
    \multirow{3}[2]{*}{MidQR($\tau_{10}$)} & Mean 10 & -0.001 & -0.005 & -0.030 & 0.058 & 0.001 & 0.044 & -0.011 & 0.022 & 0.039 & 0.017 & -0.006 & 0.008 & \multicolumn{3}{c}{0.869} \\
          & SE 10 & 0.013 & 1.339 & 0.816 & 1.492 & 0.913 & 0.677 & 0.423 & 1.151 & 0.665 & 0.510 & 0.383 & 0.382 & \multicolumn{3}{c}{0.723} \\
          & MCSE 10 & 0.00003 & 0.00144 & 0.00099 & 0.00166 & 0.00108 & 0.00078 & 0.00035 & 0.00113 & 0.00061 & 0.00037 & 0.00034 & 0.00041 & \multicolumn{3}{c}{0.00095} \\
    \midrule
    \multirow{3}[2]{*}{MidQR($\tau_{13}$)} & Mean 13 & -0.002 & -0.014 & -0.022 & 0.051 & 0.007 & 0.026 & -0.004 & 0.017 & 0.036 & 0.015 & -0.003 & 0.008 & \multicolumn{3}{c}{1.036} \\
          & SE 13 & 0.011 & 1.208 & 0.732 & 1.346 & 0.819 & 0.605 & 0.374 & 0.953 & 0.549 & 0.432 & 0.333 & 0.326 & \multicolumn{3}{c}{0.607} \\
          & MCSE 13 & 0.00002 & 0.00094 & 0.00069 & 0.00109 & 0.00075 & 0.00063 & 0.00031 & 0.00097 & 0.00052 & 0.00031 & 0.00029 & 0.00035 & \multicolumn{3}{c}{0.00080} \\
    \midrule
    \multirow{3}[2]{*}{MidQR($\tau_{16}$)} & Mean 16 & -0.004 & -0.055 & -0.030 & 0.073 & 0.012 & 0.004 & -0.003 & -0.003 & 0.051 & 0.021 & 0.001 & 0.006 & \multicolumn{3}{c}{1.284} \\
          & SE 16 & 0.014 & 1.018 & 0.642 & 1.166 & 0.726 & 0.791 & 0.482 & 0.798 & 0.459 & 0.552 & 0.397 & 0.344 & \multicolumn{3}{c}{0.632} \\
          & MCSE 16 & 0.00003 & 0.00089 & 0.00060 & 0.00093 & 0.00066 & 0.00073 & 0.00044 & 0.00095 & 0.00055 & 0.00046 & 0.00039 & 0.00039 & \multicolumn{3}{c}{0.00136} \\
    \bottomrule
    \end{tabular}%
\end{adjustbox}
  \label{tab: estimates, real_all}
\end{sidewaystable}

\begin{sidewaystable}[htbp]
  \centering
  \caption{Parameter estimates from data regarding real cyber-vulnerabilities. Only technical and contextual variables have been used as regressors.}
  \resizebox{0.8\textwidth}{!}{%
    \begin{tabular}{clrrrrrrccc}
\cmidrule{3-11}    \multicolumn{2}{c}{\multirow{2}[4]{*}{}} & \multicolumn{1}{c}{\textbf{Exposure}} & \multicolumn{2}{c}{\textbf{AV}} & \multicolumn{2}{c}{\textbf{AC}} & \multicolumn{1}{c}{\textbf{Exploit}} & \multicolumn{3}{c}{\textbf{Intercept(s)}} \\
\cmidrule{3-11}    \multicolumn{2}{c}{} &       & \multicolumn{1}{c}{\textbf{L}} & \multicolumn{1}{c}{\textbf{Q}} & \multicolumn{1}{c}{\textbf{L}} & \multicolumn{1}{c}{\textbf{Q}} &       & \textbf{1|2} & \textbf{2|3} & \textbf{3|4} \\
        \midrule
    \multirow{3}[2]{*}{OrdReg} & Mean  & -0.011 & -0.050 & 0.626 & -0.003 & 0.120 & 0.189 & \multicolumn{1}{r}{-2.610} & \multicolumn{1}{r}{0.924} & \multicolumn{1}{r}{3.086} \\
          & SE    & 0.009 & 0.468 & 0.281 & 0.279 & 0.210 & 0.218 & \multicolumn{1}{r}{0.424} & \multicolumn{1}{r}{0.408} & \multicolumn{1}{r}{0.462} \\
          & MCSE  & 0.00024 & 0.01847 & 0.01085 & 0.00871 & 0.00580 & 0.00649 & \multicolumn{1}{r}{0.01499} & \multicolumn{1}{r}{0.01408} & \multicolumn{1}{r}{0.01614} \\
    \midrule
    \midrule
    \multirow{3}[2]{*}{LinReg} & Mean  & -2.245 & -44.114 & 90.131 & -14.826 & 18.310 & 18.512 & \multicolumn{3}{c}{300.791} \\
          & SE    & 0.790 & 42.870 & 25.589 & 25.638 & 19.344 & 20.151 & \multicolumn{3}{c}{36.808} \\
          & MCSE  & 0.02076 & 1.56012 & 0.91577 & 0.70265 & 0.52643 & 0.59178 & \multicolumn{3}{c}{1.17621} \\
    \midrule
    \midrule
    \multirow{3}[2]{*}{MidQR($\tau_{1}$)} & Mean  & 0.001 & 0.045 & 0.008 & 0.010 & -0.009 & -0.004 & \multicolumn{3}{c}{0.050} \\
          & SE    & 0.020 & 2.267 & 1.322 & 0.798 & 0.618 & 0.594 & \multicolumn{3}{c}{1.316} \\
          & MCSE  & 0.00003 & 0.00096 & 0.00068 & 0.00031 & 0.00033 & 0.00033 & \multicolumn{3}{c}{0.00118} \\
    \midrule
    \multirow{3}[2]{*}{MidQR($\tau_{4}$)} & Mean  & 0.000 & 0.047 & 0.036 & 0.023 & -0.005 & -0.006 & \multicolumn{3}{c}{0.412} \\
          & SE    & 0.018 & 1.848 & 1.079 & 0.685 & 0.541 & 0.539 & \multicolumn{3}{c}{1.092} \\
          & MCSE  & 0.00003 & 0.00112 & 0.00075 & 0.00036 & 0.00033 & 0.00036 & \multicolumn{3}{c}{0.00116} \\
    \midrule
    \multirow{3}[2]{*}{MidQR($\tau_{7}$)} & Mean  & -0.001 & 0.036 & 0.038 & 0.024 & -0.005 & -0.004 & \multicolumn{3}{c}{0.672} \\
          & SE    & 0.016 & 1.573 & 0.919 & 0.626 & 0.488 & 0.491 & \multicolumn{3}{c}{0.952} \\
          & MCSE  & 0.00003 & 0.00095 & 0.00058 & 0.00030 & 0.00029 & 0.00033 & \multicolumn{3}{c}{0.00101} \\
    \midrule
    \multirow{3}[2]{*}{MidQR($\tau_{10}$)} & Mean  & -0.002 & 0.024 & 0.035 & 0.022 & -0.003 & 0.000 & \multicolumn{3}{c}{0.870} \\
          & SE    & 0.011 & 1.110 & 0.648 & 0.414 & 0.334 & 0.342 & \multicolumn{3}{c}{0.665} \\
          & MCSE  & 0.00002 & 0.00079 & 0.00047 & 0.00026 & 0.00025 & 0.00028 & \multicolumn{3}{c}{0.00081} \\
    \midrule
    \multirow{3}[2]{*}{MidQR($\tau_{13}$)} & Mean  & -0.002 & 0.013 & 0.032 & 0.018 & -0.002 & 0.000 & \multicolumn{3}{c}{1.034} \\
          & SE    & 0.010 & 0.933 & 0.545 & 0.360 & 0.291 & 0.299 & \multicolumn{3}{c}{0.567} \\
          & MCSE  & 0.00002 & 0.00061 & 0.00037 & 0.00021 & 0.00020 & 0.00021 & \multicolumn{3}{c}{0.00067} \\
    \midrule
    \multirow{3}[2]{*}{MidQR($\tau_{16}$)} & Mean  & -0.004 & 0.011 & 0.049 & 0.028 & -0.002 & 0.004 & \multicolumn{3}{c}{1.275} \\
          & SE    & 0.012 & 0.996 & 0.580 & 0.450 & 0.367 & 0.364 & \multicolumn{3}{c}{0.658} \\
          & MCSE  & 0.00004 & 0.00079 & 0.00052 & 0.00036 & 0.00035 & 0.00047 & \multicolumn{3}{c}{0.00158} \\
    \bottomrule
    \end{tabular}%
    }
 \label{tab: estimates, real_Techn}
\end{sidewaystable}

\begin{figure}
\begin{subfigure}{.45\textwidth}
  \centering
  \includegraphics[width=\linewidth]{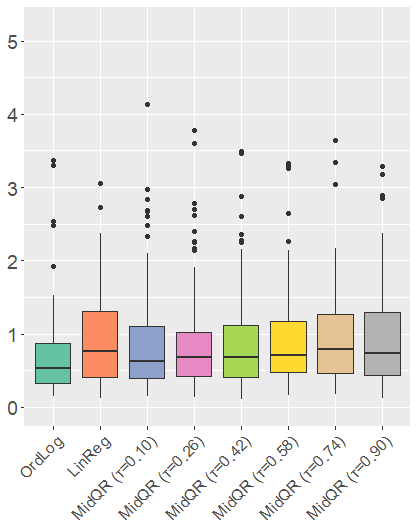}
  \caption{RGA from real data,\\ full model.}
  \label{fig: boxplot RGA real, all}
\end{subfigure}%
\begin{subfigure}{.45\textwidth}
  \centering
  \includegraphics[width=\linewidth]{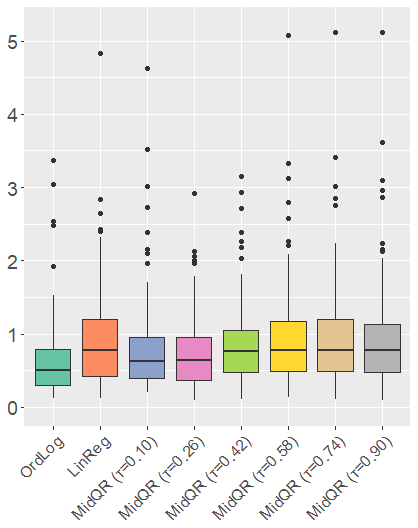}
  \caption{RGA from real data, \\ partial model.}
  \label{fig: boxplot RGA real, techn}
\end{subfigure}
\begin{subfigure}{.45\textwidth}
  \centering
  \includegraphics[width=\linewidth]{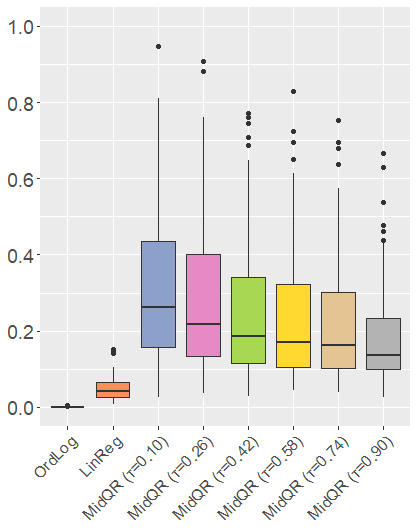}
  \caption{AGR from real data, \\ full model.}
  \label{fig: boxplot AGR real, all}
\end{subfigure}%
\begin{subfigure}{.45\textwidth}
  \centering
  \includegraphics[width=\linewidth]{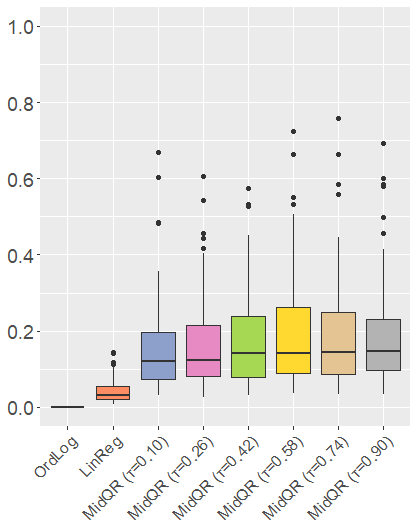}
  \caption{AGR from real data, \\ partial model.}
  \label{fig: boxplot AGR real, techn}
\end{subfigure}
\caption{Boxplots of RGA and AGR for real data. Boxplots refer, from left to right of the x-axis, to OrdLog, LinReg, and MidQR with $\tau$ taking values in $\{0.1, 0.26, 0.42, 0.58, 0.74, 0.9\}$. To improve the quality of Figure \ref{fig: boxplot AGR real, all}, the range has been restricted and excludes 9 extreme outliers for the OrdLog model and one for MidQR($\tau_{1}$).}
\label{fig: boxplots RGA-AGR real data}
\end{figure}

\begin{figure}
\begin{subfigure}{.48\textwidth}
  \centering
  \includegraphics[width=\linewidth]{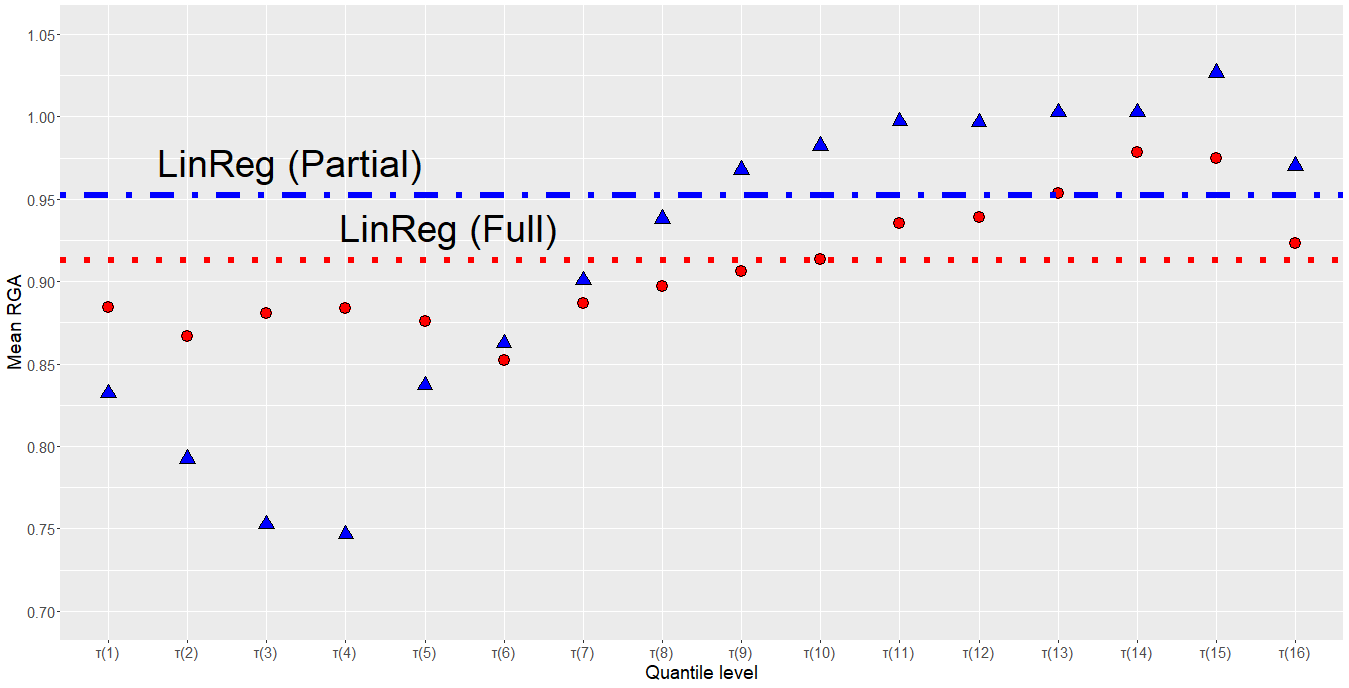}
  \caption{RGA.} 
  \label{fig: trend RGA real}
\end{subfigure}%
\hfill
\begin{subfigure}{.48\textwidth}
  \centering
  \includegraphics[width=\linewidth]{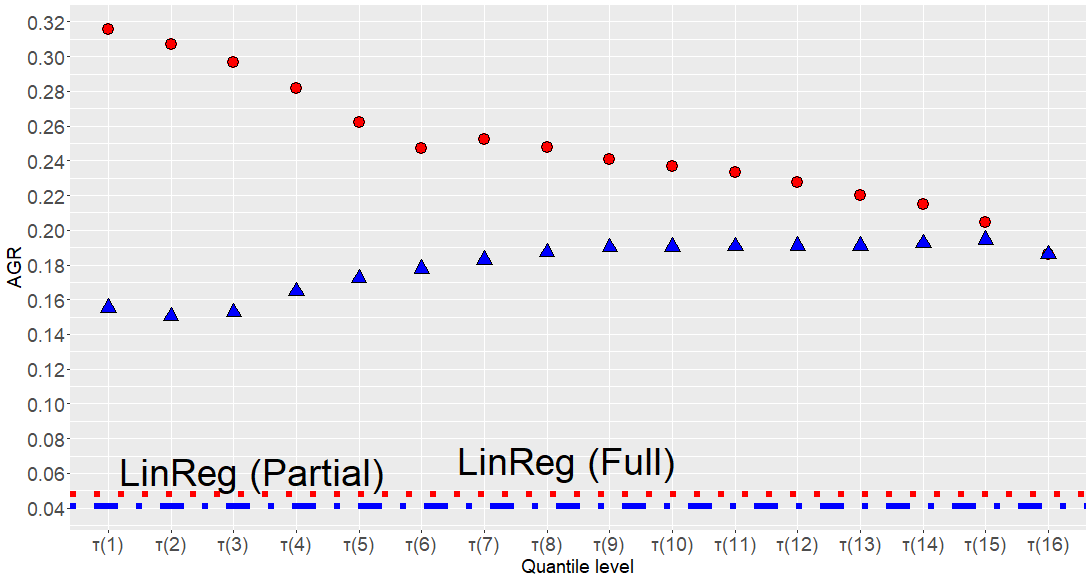}
  \caption{AGR.}
  \label{fig: trend AGR real}
\end{subfigure}
\caption{Behaviour of average RGA and AGR for real data and the $16$ quantile levels $\tau$ under consideration. Circles and triangles denote the index estimates for the full and partial models, respectively. The $y$-intercepts of the dotted and dot-dashed lines represent the value of the index from the ordered logit and the linear regression on rank-transformed variables, respectively.}
\label{fig: trends RGA-AGR}
\end{figure}

Moving to the performance indices, both RGA and AGR for all the regression models under examination are reported in Table \ref{tab: RGA-AGR real data}. In addition, we provide two graphical representations regarding the behaviour of the predictive performance at different quantile levels: the boxplots in Figure \ref{fig: boxplots RGA-AGR real data} and the plots of average RGA and AGR for all $16$ quantile levels in Figure \ref{fig: trends RGA-AGR}.


\begin{table}[htbp]
  \centering
  \caption{RGA and AGR indices from real data analysis. Columns 2--5 refer to models with the full set of regressors; columns 6--9 follow from the restriction to technical (AV, AC) and contextual (exposure, exploit) variables as regressors.}
    \begin{tabular}{lcccccccc} 
    \cmidrule{2-9} \morecmidrules 
\cmidrule{2-9}    \multirow{3}[6]{*}{} & \multicolumn{4}{c}{Full set of regressors} & \multicolumn{4}{c}{Only technical regressors} \\
\cmidrule(r){2-5} \cmidrule(r){6-9}          & \multicolumn{2}{c}{\textbf{RGA}} & \multicolumn{2}{c}{\textbf{AGR}} & \multicolumn{2}{c}{\textbf{RGA}} & \multicolumn{2}{c}{\textbf{AGR}} \\
\cmidrule(r){2-3} \cmidrule(r){4-5} \cmidrule(r){6-7} \cmidrule(r){8-9}          & \multicolumn{1}{c}{Est} & \multicolumn{1}{c}{SD} & \multicolumn{1}{c}{Est} & \multicolumn{1}{c}{SD} & \multicolumn{1}{c}{Est} & \multicolumn{1}{c}{SD} & \multicolumn{1}{c}{Est} & \multicolumn{1}{c}{SD} \\
        \midrule
    \multicolumn{1}{c}{OrdLog} & 0.688 & 0.580 & 0.361 & 1.303 & 0.662 & 0.570 & 0.000 & 0.000 \\
    \multicolumn{1}{c}{LinReg} & 0.913 & 0.641 & 0.048 & 0.031 & 0.952 & 0.739 & 0.041 & 0.029 \\
    \multicolumn{1}{c}{MidQR($\tau_{1}$)} & 0.884 & 0.726 & 0.316 & 0.213 & 0.832 & 0.721 & 0.155 & 0.119 \\
    \multicolumn{1}{c}{MidQR($\tau_{2}$)} & 0.867 & 0.709 & 0.307 & 0.209 & 0.792 & 0.681 & 0.150 & 0.115 \\
    \multicolumn{1}{c}{MidQR($\tau_{3}$)} & 0.880 & 0.728 & 0.296 & 0.203 & 0.753 & 0.607 & 0.153 & 0.116 \\
    \multicolumn{1}{c}{MidQR($\tau_{4}$)} & 0.884 & 0.735 & 0.282 & 0.195 & 0.747 & 0.523 & 0.165 & 0.118 \\
    \multicolumn{1}{c}{MidQR($\tau_{5}$)} & 0.876 & 0.701 & 0.262 & 0.186 & 0.837 & 0.614 & 0.172 & 0.121 \\
    \multicolumn{1}{c}{MidQR($\tau_{6}$)} & 0.852 & 0.672 & 0.247 & 0.177 & 0.863 & 0.607 & 0.178 & 0.126 \\
    \multicolumn{1}{c}{MidQR($\tau_{7}$)} & 0.887 & 0.710 & 0.252 & 0.186 & 0.901 & 0.635 & 0.183 & 0.131 \\
    \multicolumn{1}{c}{MidQR($\tau_{8}$)} & 0.897 & 0.708 & 0.247 & 0.183 & 0.938 & 0.708 & 0.187 & 0.137 \\
    \multicolumn{1}{c}{MidQR($\tau_{9}$)} & 0.906 & 0.694 & 0.241 & 0.179 & 0.968 & 0.748 & 0.190 & 0.139 \\
    \multicolumn{1}{c}{MidQR($\tau_{10}$)} & 0.914 & 0.683 & 0.237 & 0.176 & 0.983 & 0.776 & 0.191 & 0.140 \\
    \multicolumn{1}{c}{MidQR($\tau_{11}$)} & 0.936 & 0.696 & 0.233 & 0.174 & 0.998 & 0.781 & 0.191 & 0.142 \\
    \multicolumn{1}{c}{MidQR($\tau_{12}$)} & 0.939 & 0.680 & 0.227 & 0.173 & 0.997 & 0.786 & 0.191 & 0.143 \\
    \multicolumn{1}{c}{MidQR($\tau_{13}$)} & 0.954 & 0.666 & 0.220 & 0.164 & 1.003 & 0.791 & 0.191 & 0.142 \\
    \multicolumn{1}{c}{MidQR($\tau_{14}$)} & 0.978 & 0.675 & 0.215 & 0.157 & 1.003 & 0.792 & 0.192 & 0.142 \\
    \multicolumn{1}{c}{MidQR($\tau_{15}$)} & 0.975 & 0.679 & 0.205 & 0.150 & 1.027 & 0.790 & 0.195 & 0.141 \\
    \multicolumn{1}{c}{MidQR($\tau_{16}$)} & 0.923 & 0.675 & 0.186 & 0.131 & 0.970 & 0.797 & 0.186 & 0.134 \\
    Self  & 6.275 & 1.095 & 6.275 & 1.095 & 6.327 & 1.123 & 6.327 & 1.123 \\
    \bottomrule
    \end{tabular}
  \label{tab: RGA-AGR real data}
\end{table}

In order to investigate the robustness of the analysis according to the aforementioned settings, we conducted parallel analyses with different partitionings ($n_{tr} = 574$ and $n_{test} = 140$), a different number of iterations, or scaling of the numerical regressor. The results and overall performance in the different scenarios are similar to those we have presented above, revealing a satisfactory robustness of the proposed approach. 

\section{Discussion}
\label{sec: discussion}
In line with the search for flexibility, interpretability, and robustness in cyber-risk assessments, a quantile-based approach can extract relevant information beyond means to examine rare events, which is a primary need for the continuity of a network or critical infrastructure. The AGR index lets us evaluate predictive performance without relying on a quantitative structure for the ordinal responses. Here, we discuss the outcomes of the analysis of synthetic and real data. 

\paragraph{AGR as an appropriate measure of predictive accuracy}  
From simulations, we see that the data-generating models are generally associated with a higher AGR value, while their RGA is often worse than other models (see Figures \ref{fig: boxplots RGA-AGR simulations, k=4}, \ref{fig: boxplots RGA-AGR simulations, k=3}, and \ref{fig: boxplots RGA-AGR simulations, k=6 and k=8}). 
It is plausible that the specific model underlying the data generation process provides better predictive performance compared to other models. This criterion identifies AGR as a more appropriate performance index for our purposes since it better distinguishes the data-generating model in terms of predictive capacity, as is manifest from the above-mentioned figures. 

In addition, AGR enjoys the invariance property under sub-sampling, as discussed in Section \ref{sec: our proposal}, which is desirable since the measure is not affected by other (possibly unknown) vulnerabilities. In this way, we can better prioritise the vulnerabilities under consideration without incurring order reversal due to new vulnerabilities not previously detected. From a different perspective, such new information may be needed to update individual priority ratings and adapt to the dynamic behaviour of cyber-space, as is discussed in the following paragraph. 

\paragraph{MidQR and probabilistic risk modelling} 
We already pointed out the distinction between cyber-incidents and cyber-vulnerability. Recalling that the analysis in \citet{Giudici2021} focuses on the former, the comparison of the regression models that we have carried out is purely methodological, and the tests we conducted on cyber-vulnerability data set a common ground to compare the characteristics of the methods in terms of RGA and AGR indices. By the same token, the rank transform has been used to enhance the comparability of the responses produced by the two models. 

In this regard, while rankings are the primary outcome of LinReg, mid-quantile models produce cumulative probability estimates for ordinal responses. A potential extension of this research is the comparison of different conditional (mid-)probabilities extracted from mid-quantile methods obtained with different sets of regressors; the information divergence between such distributions, e.g., through entropy-based methods, can support the quantification of the information content provided by the vulnerability's characteristics. In this way, our proposal can support the search for new models for cyber-risk analysis based on probability and impact \citep{allodi2017security}. 

While the present work uses Tenable's VPR for the analysis, each decision-maker can customise the model (as well as the quantile level), adapt it in time to get new estimates and quantile effects, or compare different risk factors derived from different criteria in terms of predictive power. This opportunity stimulates further studies to take advantage of probability estimates from mid-quantile methods in specific scenarios or case studies. Indeed, networks of connected organisations could carry out the analysis using their own threat assessment as the response variable; therefore, such probability estimates could help conduct risk analysis in conjunction with Bayes update rules and graphical models, e.g., Bayesian networks \citep{shin2015development}, providing an alternative to the assignment of standard values for probabilities starting from qualitative experts' opinions. We also stress that the proposed approach can be extended to quantitative response variables too; indeed, we can choose a different set of regressors related to cyber-vulnerabilities' characteristics and severity, considering the frequency of related cyber-incidents as a response variable, if available. In this way, the fitted mid-cumulative distribution functions could represent a robust alternative to estimating or predicting the number of cyber-incidents or cyber-intrusions \citep{leslie2018statistical}.

\paragraph{Real and synthetic data}  

Referring to Table \ref{tab: RGA-AGR real data}, two different models are considered: the full one (all the relevant variables in the dataset derived from Table \ref{tab: data structure} are involved) and a restricted one, where the ``CIA'' components of attack vectors are excluded. This choice is driven by a better understanding of the role of the CVSS impact dimensions in vulnerability prioritisation and cyber threat analysis \citep{allodi2014comparing,allodi2017security}. Table \ref{tab: RGA-AGR real data} suggests that different regression models provide different information regarding the role of the CIA attributes, where OrdLog generates larger deviations (outliers) with high accuracy that seriously affect the average accuracy performance; clearly, quantile-based indices depicted in Figure \ref{fig: boxplots RGA-AGR real data} are more robust with regard to such anomalies. Furthermore, the two models show different behaviours at varying quantile levels, as exhibited in Figure \ref{fig: trends RGA-AGR}. 
 
By comparing the full and partial models, we observe that AGR leads to higher discrimination than RGA does. Formally, let us consider the ratios
\begin{equation} 
\varrho_{\mathrm{RGA}}:=\frac{\overline{\mathrm{RGA}_{\mathrm{tech}}}}{\overline{\mathrm{RGA}_{\mathrm{full}}}},\quad \varrho_{\mathrm{AGR}}:=\frac{\overline{\mathrm{AGR}_{\mathrm{tech}}}}{\overline{\mathrm{AGR}_{\mathrm{full}}}}
\end{equation}  
of the average values of RGA and AGR evaluated for the technical and full models, respectively. For the LinReg model, AGR leads to higher discrimination than RGA does ($\varrho_{\mathrm{RGA}}=1.043$ and $\varrho_{\mathrm{AGR}}=0.862$). Focusing on MidQR, we also see that AGR is more sensitive than RGA with respect to the choice of the quantile level in terms of model discrimination. Indeed, $\varrho_{\mathrm{RGA}}\in [0.845;1.076]$, while $\varrho_{\mathrm{AGR}}\in [0.490;1.002]$, and $\varrho_{\mathrm{AGR}}<0.8$ for quantile levels $\tau_{1}$ to $\tau_{9}$. In fact, $\varrho_{\mathrm{AGR}}$ tends to increase with the quantile level, which suggests a non-trivial contribution of the CIA attributes in combination with information about exposure or exploits, which also depends on the choice of the quantile level. 

While the LinReg and MidQR models considered in this work are comparable in terms of RGA performance on real data, using AGR, we can see that OrdLog performs poorly since the predicted values are restricted to the set $\{1,\dots,k\}$. When the dataset has low variability, the estimated values collapse to a typical value, which contains no information and drastically reduces predictive performance. This also suggests a severe deviation from the OrdLog model assumptions in the present cyber-vulnerability dataset. 

Another indirect test of the deviation of real data from the OrdLog model comes from the relative magnitude of RGA and AGR. In Tables \ref{tab: estimate, k=4 uniform}--
\ref{tab: estimates, k=8 non-uniform}, which refer to data simulated starting from the ordered logit model, AGR is comparable with RGA (i.e., with the same order of magnitude), and at low values of $k$, especially at $k=3$, AGR is larger than RGA when we focus on MidQR and the data-generating model. On the other hand, real data lead to different behaviour: calculating the ratios $\mathrm{AGR}/\mathrm{RGA}$ within each iteration, their median lies in $[0.218, 0.402]$ for the $16$ quantile levels in the full model and $[0.174, 0.213]$ in the partial model; looking at the ratios $\overline{\mathrm{AGR}}/\overline{\mathrm{RGA}}$ of the mean values shown in Table \ref{tab: RGA-AGR real data}, they range in $[0.201, 0.357]$ for the full model and in $[0.187, 0.221]$ for the partial model. These ratios are useful as an additional check of the deviation from the OrdLog model used in simulations, AGR and RGA indices for the same model should not be compared, as they measure different performance aspects of a given model. 

\paragraph{Dependence of the MidQR performance on $k$} 
MidQR performs better when the number of levels $k$ of the response variable is small (less than $6$), as can be seen comparing Figures \ref{fig: boxplots RGA-AGR simulations, k=4}-\ref{fig: boxplots RGA-AGR simulations, k=3} with Figure \ref{fig: boxplots RGA-AGR simulations, k=6 and k=8}. In the latter, AGR highlights a divergence between the data-generating model (OrdLog) and alternative models (LinReg or MidQR); on the other hand, RGA returns a performance comparable to that of LinReg and MidQR.
    
\paragraph{SE of the estimates} 
As remarked in the previous section, an arbitrary choice of the quantile level may lead to overestimating the parameter SE through the kernel approach adopted in \citep{Geraci2022} and based on \citep{Li2008}; this is confirmed by the outputs of the simulations. When this overestimation happens, the remaining indices (i.e., the regular SE and the MCSE) provide a more informative picture of the sampling distribution. 

\paragraph{Implications for cyber-threat intelligence and secure information disclosure} 

As a practical consequence of the observations in the last paragraphs, we draw attention to the information the individual decision-maker has, uses, and communicates about cyber-risk. 

Agencies such as NIST share their evaluation through dedicated information channels; however, this information can also be acquired by potential attackers, who can use it to prioritise their own objectives. Indeed, resources are also needed by attackers (e.g., costs for exploit acquisition, time and effort for detection of vulnerable hosts, integration of multiple components to avoid countermeasures), and information on risk factors from different organisations can be useful to suggest relevant criticalities. 

Our proposal addresses this issue in two ways: first, as already recalled, MidQR enhances robustness against violations of assumptions in parametric methods and allows for the analysis of different types of explanatory or response variables; this makes MidQR suited to compare models with different sets of explanatory variables and then choose an appropriate trade-off between predictive ranking accuracy and limited information to be shared. The second contribution involves the invariance property of the AGR index, which avoids inconsistency in rankings obtained from different sets of cyber-vulnerabilities in the sense of Example \ref{exa: non-invariance}; this reduces the need to share information on relevant cyber vulnerabilities to achieve a given value of accuracy in rank estimation. 

These observations are mainly related to cybersecurity data and their usefulness for distinct decision-making stages, which led us to select the databases described in Section \ref{sec: data sources}. Information granularity in data from cyber-incidents does not often suffice to extract useful insights into the current threats. This leads to data aggregation and censoring that could not allow cybersecurity operational experts to prioritise the current vulnerabilities, as is the case in the classification of attack techniques reported in \citep{Giudici2021}, where multiple types of attacks are grouped together (e.g., SQL injection is a particular attack model upon which malware can be based, and malware can exploit one or more 0-days). Similarly, the use of ordered logit or other GLMs is a well-established approach to carrying out inference about probabilities, even in the cyber-risk domain \citep{mukhopadhyay2019cyber}, but the present analysis has shown that it is not suited to the collected cyber-vulnerability data. However, this should be interpreted as complementarity between the analyses on cyber-incidents, and the present one: they serve different phases (strategic, tactic, or operative) of a process with a common objective, and each phase should identify appropriate data for its scope.



\section{Conclusion and Future Work}
\label{sec: conclusion} 

This work investigated statistical modelling for threat intelligence, with particular attention to the information resources regarding cyber-vulnerabilities. Being fixing resource-expensive, decision-makers have to allocate their resources based on their current state of knowledge and their risk perception. The statistical model and the index proposed for cyber-vulnerability assessment complement other approaches developed in the cyber-risk literature. These models are not mutually exclusive and could be considered in parallel to highlight distinct aspects of relevance to decision-makers. 

The actual realisation of cyber-attacks relies on several information sources that can enhance or inhibit them. It is plausible that indirect access to information plays a more important role than expected: along with limited data disclosure and underreporting, even prioritisation data communicated by organisations to prevent cyber-incidents can guide cyber-attackers, as discussed in Section \ref{sec: discussion}. The present work opens the way to further applications supporting secure information disclosure on cyber-vulnerabilities, since the advantages of the framework discussed in the previous sections can highlight the effects of both information sources (in terms of available regressors) and cyber-risk perception or severity assessments (e.g., a suitable data-generating model). A more accurate evaluation of such effects is a necessary premise to avoid the indirect and unintended communication of information. 
 
A deeper investigation is needed for the emergence of multiple prioritisations due to different decision criteria and uncertainty sources, which may occur when different experts or organisations conduct separate analyses based on their own choices for response and explanatory variables. Various approaches could be explored to formalise compatibility conditions for ordinal structures under uncertainty \citep{Angelelli2024} in continuity with the arguments that led to the AGR index in Section \ref{subsec: a new performance index for cyber-risk estimation}. A dedicated study to identify information-theoretic, fuzzy, or relational criteria to encompass and quantify specific uncertainty sources in cyber-space could support individuals or groups in contextualising risk assessment about shared digital resources. 

Despite the generality of the methodology, a limitation of this work is that it does not explicitly consider context-specific data that could affect cyber-vulnerability prioritisation. Risk factors may vary due to internal priorities in the organisation and the evolution of the overall digital system (new products, legislation). Patterns extracted within Tenable's VPR processing contain information about risks posed by cyber-threats, but contextual factors should also be explored when adapting this analysis to specific case studies or operational scenarios, including governance requirements, tools for the development of secure digital products \citep{Baldassarre2020}, privacy \citep{baldassarre2020integrating}, and behavioural factors that can influence the perception of the exploitability of a cyber-vulnerability. Future work will explore complementary approaches for estimating behavioural latent traits, including Bayesian methods, and connecting them to relevant parameters in risk assessment (e.g., the choice of the quantile level). 
These factors require specific measurement models and evaluation methods, and, in line with the adoption of graphical methods in cyber-risk assessment, structural equation models \citep{woods2021sok} could be a valid option to extend our research directions into the study of behavioural risk perception.  

\section*{Abbreviations}
AC (Access Complexity), AGR	(Agreement of Grounded Ranking), AV (Access Vector), CDF	(Cumulative Distribution Function), CIA	(Confidentiality, Integrity, and Availability), CSIRT (Computer Security Incident Response Team), CVE (Common Vulnerability Exposure), CVSS (Common Vulnerability Scoring System), FAIR (Factor Analysis of Information Risk), GDPR (General Data Protection Regulation), GLM (Generalised Linear Model), ICT (Information and Communication Technology), 
IoT (Internet-of-Things), 
MCSE (Monte Carlo Standard Error), NIST (National Institute of Standards and Technology), NVD (National Vulnerability Database), QQ (Quantile-Quantile), RGA (Rank Graduation Accuracy), SE (standard error), VaR (Value-at-Risk), VPR (Vulnerability Priority Rating). 

\section*{Data availability}
The sources of data collected for the analysis are publicly available. Scripts are available upon request. 

\section*{Acknowledgements}
Mario Angelelli is a member of the Istituto Nazionale di Alta Matematica (INdAM-GNSAGA). Serena Arima acknowledges the financial support provided by the MiuR-PRIN Grant No 2022Z85NCT (Violence against women: modelling misreported information in social data, PI: Serena Arima). Christian Catalano acknowledges the publication was produced with the co-funding of the European Union - Next Generation EU: NRRP Initiative, Mission 4, Component 2, Investment 1.3 - Next Generation EU (PE0000014 - ``SEcurity and Rights In the CyberSpace - SERICS'' - CUP: H93C22000620001). 





\bibliographystyle{apalike}
\bibliography{references}

\end{document}